\documentclass[conference,10pt]{IEEEtran}
\IEEEoverridecommandlockouts

\usepackage[utf8]{inputenc}
\usepackage[T1]{fontenc}
\usepackage{cite}
\usepackage{amsmath,amssymb}
\usepackage{graphicx}
\usepackage{booktabs}
\usepackage{multirow}
\usepackage{xcolor}
\usepackage{url}
\usepackage[hidelinks]{hyperref}
\hypersetup{
  pdftitle={Thinking in Graphs: Neo4j-Backed Exploit-Chain Reasoning for LLM Penetration-Testing Agents},
  pdfauthor={Joas Antonio dos Santos},
  pdfsubject={attack graphs; Neo4j; LLM pentest agents; exploit chains; MCP; Active Directory},
  pdfkeywords={attack graph, Neo4j, exploit chain, LLM agent, penetration testing, Model Context Protocol, Cypher, Active Directory, chain severity}
}
\usepackage{listings}
\usepackage{tikz}
\usetikzlibrary{shapes.geometric, arrows.meta, positioning, fit, backgrounds, calc, patterns, decorations.markings}
\usepackage{pgfplots}
\pgfplotsset{compat=1.18}
\usepackage{float}
\usepackage{enumitem}
\usepackage{placeins}
\usepackage{balance}
\usepackage{array}
\usepackage{caption}

\lstdefinelanguage{Cypher}{
  morekeywords={MATCH, OPTIONAL, WHERE, RETURN, WITH, CALL, YIELD, CREATE, MERGE, CONSTRAINT, INDEX, FOR, REQUIRE, UNIQUE, ON, AS, ORDER, BY, LIMIT, DISTINCT, collect, shortestPath, length, nodes, size, round, coalesce, count, CASE, WHEN, THEN, ELSE, END, IF, NOT, EXISTS, left},
  sensitive=true,
  morecomment=[l]{//},
  morestring=[b]',
  morestring=[b]",
  keywordstyle=\color{blue!70!black}\bfseries,
  commentstyle=\color{green!45!black}\itshape,
  stringstyle=\color{red!55!black},
}
\lstdefinelanguage{Json}{
  morestring=[b]",
  keywordstyle=\color{blue!70!black},
  morekeywords={true,false,null},
  commentstyle=\color{gray}\itshape,
  stringstyle=\color{red!50!black},
}
\lstdefinelanguage{Toml}{
  morecomment=[l]{\#},
  morestring=[b]",
  keywordstyle=\color{blue!70!black},
  commentstyle=\color{green!45!black}\itshape,
  stringstyle=\color{red!55!black},
}
\lstdefinelanguage{Shell}{
  morekeywords={claude, codex, docker, uvx, neo4j, export, mcp, add},
  sensitive=true,
  morecomment=[l]{\#},
  morestring=[b]",
  keywordstyle=\color{blue!70!black}\bfseries,
  commentstyle=\color{green!45!black}\itshape,
  stringstyle=\color{red!55!black},
}

\definecolor{critcolor}{RGB}{229,72,77}
\definecolor{highcolor}{RGB}{247,107,21}
\definecolor{medcolor}{RGB}{217,160,0}
\definecolor{lowcolor}{RGB}{62,123,250}
\definecolor{infocolor}{RGB}{139,134,152}
\definecolor{graphgreen}{RGB}{16,150,110}
\definecolor{directred}{RGB}{239,68,68}
\definecolor{graphblue}{RGB}{59,130,246}
\definecolor{graphpurple}{RGB}{124,92,236}
\definecolor{nodefind}{RGB}{245,179,1}
\definecolor{nodestep}{RGB}{124,92,236}
\definecolor{nodecred}{RGB}{16,150,110}
\definecolor{nodeasset}{RGB}{229,72,77}

\newcommand{\cmark}{\textcolor{green!60!black}{\checkmark}}
\newcommand{\pmark}{\textcolor{orange!80!black}{$\sim$}}
\newcommand{\xmark}{\textcolor{red!60!black}{---}}

\begin{document}

\title{Thinking in Graphs: Neo4j-Backed\\Exploit-Chain Reasoning for LLM\\Penetration-Testing Agents}

\author{\IEEEauthorblockN{Joas Antonio dos Santos}
\IEEEauthorblockA{\textit{Independent Researcher --- AI and Offensive Security} \\
S\~ao Paulo, SP, Brazil \\
joas.santos@redteamleaders.com \\
ORCID: 0009-0003-1772-9826}
}

\maketitle

\begin{abstract}
Large language model (LLM) coding agents such as Claude Code and OpenAI Codex can already drive individual penetration-testing tools, but they reason about an engagement as a flat list of findings held in a context window. Attackers, by contrast, think in graphs: isolated low-severity exposures become catastrophic once chained through credential capture, privilege escalation, and lateral movement. This gap feeds a common belief that AI agents cannot truly \emph{chain} exploitation end to end---a belief we argue describes the flat context window, not the model: give the agent a persistent attack graph and chaining becomes a deterministic database operation rather than an emergent hope. We present \textsc{ChainGraph}, a lightweight pattern that externalizes exploit-chain reasoning into a Neo4j property graph connected directly to an off-the-shelf coding agent via the Model Context Protocol (MCP)---no bespoke harness required. Findings, attack steps, credentials, assets, and tamper-evident evidence receipts are stored as typed nodes with pre/post-condition edges (\texttt{ENABLES}, \texttt{REQUIRES}, \texttt{YIELDS}, \texttt{ESCALATES\_TO}, \texttt{PIVOTS\_TO}); JSON and Markdown artifacts serve as the documentary/evidential memory while Neo4j serves as the relational reasoning memory. We formalize five operator questions as reusable Cypher and Graph Data Science (GDS) queries---next-step inference, entry-to-crown-jewel pathing, combined-severity promotion, evidence provenance, and chokepoint identification---and introduce a \emph{chain severity} score that is distinct from per-finding local severity. On a synthetic multi-forest Active Directory engagement modeled on GOAD and loaded into Neo4j~5.26 with GDS~2.13, the graph promotes three findings rated Medium in isolation to Critical once their path to the forest root is materialized, and betweenness centrality identifies the PKINIT authentication and DCSync steps as the highest-value chain break-points. Returning a single eight-hop attack path as a graph query result rather than re-feeding the raw engagement corpus reduced the reasoning payload by roughly $98\%$ on a character-based token proxy. We give a complete, reproducible configuration---Docker, schema, queries, MCP wiring for both Claude Code and Codex, and a PostToolUse evidence-capture hook---and discuss the security, calibration, and dual-use implications of giving an autonomous agent a persistent attack graph. We frame all quantitative observations as illustrative single-run measurements, not a controlled benchmark.
\end{abstract}


\section{Introduction}\label{sec:intro}

In 2015, John Lambert captured the asymmetry at the heart of network defense in a single sentence: ``Defenders think in lists; attackers think in graphs. As long as this is true, attackers win''~\cite{lambert2015defenders}. A vulnerability scanner emits a list. A compliance report emits a list. A ticketing queue emits a list. The adversary, however, does not care about any single finding---they care about the \emph{path}: the sequence of individually unremarkable conditions that, composed, carries them from an unauthenticated network position to control of the enterprise.

The recent wave of LLM-driven penetration-testing agents inherits the defender's list-shaped worldview almost by accident. PentestGPT~\cite{deng2024pentestgpt}, AutoAttacker~\cite{xu2024autoattacker}, Incalmo~\cite{singer2025incalmo}, VulnBot~\cite{kong2025vulnbot}, and autonomous Active Directory (AD) agents~\cite{happe2025ad} all keep their working model of the engagement inside the LLM context window---a flat, linear transcript of observations and tool outputs. This representation has three structural weaknesses. First, it degrades with length: transformer models attend unevenly to long contexts, and relevant facts buried in the middle are effectively lost~\cite{liu2024lost}. Second, it conflates \emph{storage} with \emph{reasoning}: every question about the engagement---``does finding X reach a domain admin?''---requires re-reading the entire transcript, because there is no index over relationships. Third, it provides no stable, queryable notion of a \emph{chain}: the agent must re-derive, token by token, that a Medium-severity LLMNR exposure feeds a crackable hash, which feeds an AD~CS misconfiguration, which yields a forged domain-admin certificate.

Consider the running example we return to throughout the paper. An agent captures an LLMNR response, cracks the resulting hash to a domain-user password, finds an Active Directory Certificate Services template that lets any enrollee request a certificate for an arbitrary principal, forges a domain-admin certificate, authenticates with it, and runs DCSync. In a list-shaped report these are five findings: one Medium (LLMNR), one Medium (weak password), one High (the template), and two operational notes. In a graph they are one \emph{path} from an unauthenticated position to the forest root---and the first Medium is the match that lights the whole chain. No LLM reasoning over a flat transcript reliably recovers that path on demand; a \texttt{shortestPath} query recovers it in milliseconds and returns only the answer.

The classical security literature solved the representation problem decades ago. Attack graphs---formalized by Phillips and Swiler~\cite{phillips1998graph}, automated by Sheyner et~al.~\cite{sheyner2002automated}, scaled by MulVAL~\cite{ou2005mulval} and NetSPA~\cite{ingols2006practical}, and surveyed exhaustively since~\cite{kaynar2016taxonomy,lallie2020review}---model reachability as a directed graph of pre- and post-conditions and compute security metrics over \emph{paths}, not nodes~\cite{wang2008probabilistic,homer2013aggregating}. BloodHound~\cite{robbins2016bloodhound} brought exactly this idea to AD, storing identities, hosts, and access-control relationships in Neo4j and answering ``six degrees of Domain Admin'' with a graph query. The missing piece is not the graph; it is the \emph{wiring} between the graph and the modern LLM agent that now does the hands-on-keyboard work.

\textbf{The paradigm we break.} A common objection to autonomous offensive AI holds that language models cannot truly \emph{chain} exploitation: they can drive a single tool, but they lose the thread across a multi-step campaign, declare victory on narrative plausibility, and never assemble isolated findings into an end-to-end path from foothold to domain control. This skepticism is not baseless---it accurately describes the \emph{flat-context} agent, and the pentest literature documents exactly these failures: PentestGPT losing track over long interactions~\cite{deng2024pentestgpt}, AD agents going ``down rabbit holes''~\cite{happe2025ad}. But the limitation is an artifact of the \emph{representation}, not of the model's capability. The evidence that models can chain is already in: teams of LLM agents exploit real zero-day vulnerabilities that single agents cannot~\cite{zhu2024teams}, and systems with an explicit state layer plan multi-host campaigns~\cite{singer2025incalmo}. Our claim is sharper: give the agent a persistent attack graph as external reasoning memory, and chaining stops being an emergent hope and becomes a \emph{database operation}---the eight-hop path in our case study (Section~\ref{sec:case}) is returned by a single \texttt{shortestPath} query, deterministically, every time. The paradigm ``AI can't chain exploitation'' is really ``a context window can't hold a chain.'' Replace the context window with a graph, and the chain is always there to be walked.

This paper supplies that wiring. We argue that an LLM pentest agent should treat a graph database not as an output format but as its external reasoning memory, and that this can be achieved today with an unmodified coding agent---Claude Code~\cite{claudecodedocs} or OpenAI Codex~\cite{codexdocs}---connected to Neo4j through the Model Context Protocol (MCP)~\cite{anthropic2024mcp,mcpspec,neo4jmcp}. We call the pattern \textsc{ChainGraph}. Our contributions are:

\begin{enumerate}[leftmargin=*,nosep]
\item A \textbf{dual-memory architecture} (Section~\ref{sec:arch}) that separates documentary/evidential memory (versioned JSON and Markdown with content-hashed evidence receipts) from relational reasoning memory (a Neo4j property graph), and a \textbf{typed schema} (Section~\ref{sec:schema}) whose edges encode attack-step pre/post-conditions.
\item A \textbf{reasoning layer} (Section~\ref{sec:reasoning}) expressing five concrete operator questions as reusable Cypher/GDS queries, and a \textbf{chain-severity model} that promotes findings by the paths they unlock rather than by their class in isolation.
\item A \textbf{zero-harness integration recipe} (Section~\ref{sec:integration}): step-by-step Docker, schema, MCP, permission, and evidence-capture-hook configuration for Claude Code and Codex, with read/write tool separation enforced at the protocol boundary.
\item A \textbf{reproducible case study} (Section~\ref{sec:case}) on a synthetic multi-forest AD engagement modeled on GOAD~\cite{goad}, loaded into a live Neo4j~5.26 + GDS~2.13 instance, demonstrating next-step inference, entry-to-crown-jewel pathing, severity promotion, and chokepoint analysis on real query output.
\item An \textbf{efficiency and calibration analysis} (Section~\ref{sec:eval}) measuring the reasoning-payload reduction from graph-scoped retrieval versus flat-context re-reading, and a discussion (Sections~\ref{sec:security}--\ref{sec:limits}) of the security, dual-use, and calibration implications.
\end{enumerate}

We are deliberate about scope: the quantitative figures are single-run measurements on a synthetic lab, offered to motivate and illustrate the architecture, not to establish statistically powered benchmarks. Section~\ref{sec:limits} states the threats to validity plainly.

\section{Background and Related Work}\label{sec:related}

\subsection{Attack Graphs and Path-Based Security Metrics}
The attack-graph model represents a network as nodes (system states, conditions, or privileges) and edges (atomic attacker actions that transform one state into another). Phillips and Swiler~\cite{phillips1998graph} introduced the graph-based formulation; Sheyner et~al.~\cite{sheyner2002automated} generated and analyzed such graphs automatically via model checking; MulVAL~\cite{ou2005mulval} recast generation as Datalog logic programming for scalability, and NetSPA~\cite{ingols2006practical} pushed to enterprise scale. Because attack graphs are explicitly path-structured, they support \emph{path-based} security metrics: probabilistic likelihood of compromise~\cite{wang2008probabilistic}, aggregated vulnerability metrics across an enterprise~\cite{homer2013aggregating}, and hierarchical complexity management~\cite{noel2004managing}. Kaynar~\cite{kaynar2016taxonomy} and Lallie et~al.~\cite{lallie2020review} survey the generation and visual-syntax design space. Our chain-severity model (Section~\ref{sec:severity}) is a deliberately lightweight, agent-facing descendant of these metrics, computed on demand over an incrementally built graph rather than a pre-enumerated state space.

\subsection{Graphs, Knowledge Graphs, and LLMs}
A parallel line of work connects LLMs to graph-structured knowledge. GraphRAG~\cite{edge2024graphrag} builds a knowledge graph from a corpus, detects communities, and retrieves hierarchical summaries to answer global queries that flat vector RAG cannot; two 2024--2025 surveys map the design space~\cite{peng2024graphragsurvey,han2025graphrag}. The retrieval-as-reasoning thread is closer to our use: Think-on-Graph~\cite{sun2024thinkongraph} has the LLM walk a knowledge graph beam-search style, treating each hop as an explicit reasoning step, and G-Retriever~\cite{he2024gretriever} retrieves a relevant subgraph for graph question answering---both mirror how our agent traverses the attack lattice rather than re-reading text. HippoRAG~\cite{gutierrez2024hipporag} adds Personalized PageRank over a knowledge graph as a long-term memory index, and LightRAG~\cite{guo2024lightrag} uses a dual-level graph index for cheap incremental retrieval. Pan et~al.~\cite{pan2024unifying} give the roadmap for unifying LLMs and knowledge graphs, distinguishing KG-enhanced LLMs (our case) from LLM-augmented KGs. Crucially for deployment, LLMs translate natural language into Cypher with increasing fidelity---Text2Cypher~\cite{ozsoy2024text2cypher} (from the Neo4j team), CypherBench~\cite{feng2024cypherbench} measuring retrieval precision over full-scale graphs, and SyntheT2C~\cite{zhong2025synthet2c} generating fine-tuning data---so the agent can \emph{author} the queries it needs. Cypher's formal property-graph semantics~\cite{francis2018cypher} keep those queries deterministic and auditable, a property a generative answer lacks. ChainGraph applies this stack to a security-reasoning graph: the ``documents'' are findings and the ``summaries'' are paths.

\subsection{Agent Memory}
The flat-context limitation~\cite{liu2024lost} has driven a wave of external-memory architectures: MemGPT~\cite{packer2023memgpt} paginates memory like an operating system; Mem0~\cite{chhikara2025mem0} and Zep/Graphiti~\cite{rasmussen2025zep} maintain scalable, in part graph-structured, long-term memory for agents. ChainGraph is memory specialized for one domain: the state of an offensive engagement, where the relationships (what enables what, what yields which credential) are the payload.

\subsection{Graphs for Security + LLMs}
Recent work fuses the two threads. AttacKG+~\cite{zhang2025attackgplus} and RAG-based attack-graph generation~\cite{prapty2024ragattackgraph} use LLMs to \emph{build} attack/technique graphs from threat intelligence; ATAG~\cite{gandhi2025atag} models threats to AI-agent applications with attack graphs; CRAKEN~\cite{shao2025craken} injects knowledge-based execution into a cyber LLM agent. Closest to our motivation, SENTINEL-RL~\cite{vallabhaneni2026sentinel} offloads topological reasoning from LLM agents to a graph in a defensive SOC setting---evidence that externalizing graph reasoning helps agents on both sides of the fence. ChainGraph differs in targeting the \emph{offensive engagement loop} of an unmodified coding agent and in being deployable today via MCP with no custom runtime.

\subsection{LLM Penetration-Testing Agents}
PentestGPT~\cite{deng2024pentestgpt} formalized the reason/generate/parse pipeline and reported, at USENIX Security 2024, that LLMs lose track of the engagement over long interactions---exactly the flat-context failure we externalize. AutoAttacker~\cite{xu2024autoattacker} automated post-breach hands-on-keyboard actions; Zhu et~al.~\cite{zhu2024teams} showed teams of LLM agents exploiting real vulnerabilities that single agents could not. Incalmo~\cite{singer2025incalmo} is the closest in spirit on the state question: it introduces a high-level attack abstraction layer and explicit \emph{state services} between the LLM and the network, letting the model express intent while a deterministic layer tracks hosts and credentials---ChainGraph can be read as supplying that state service as a queryable graph over commodity infrastructure. PentestAgent~\cite{shen2025pentestagent} pairs multiple LLM agents with retrieval-augmented generation for automated testing, and VulnBot~\cite{kong2025vulnbot} coordinates multiple agents over a penetration task graph, and Happe and Cito~\cite{happe2025ad} demonstrated autonomous assumed-breach AD testing, observing models going ``down rabbit holes'' for want of a global view. Our own prior NeuroSploit/GOAD benchmark~\cite{barbosa2026neurosploit} showed structured orchestration with belief tracking beats unstructured LLM usage for AD. The common thread is that \emph{state must live outside the model}; ChainGraph makes that state a first-class, queryable graph and shows it needs no bespoke harness, only MCP. The underlying agent loop is ReAct-style reason-and-act~\cite{yao2023react}.

\subsection{Commercial Autonomous Pentest Agents}
The strongest existence proof that LLM agents \emph{do} chain exploitation comes not from the lab but from shipping commercial products, and two in particular bear directly on our thesis. HackerSec's \textbf{Yaga}~\cite{hackersec2026yaga} is an in-house offensive-AI pentest agent that, by its vendor's account, performs technical reconnaissance, interprets context, and executes real in-scope exploits in a \emph{chained} rather than isolated-check fashion---built on a multi-agent architecture with tactical RAG, stigmergic planning, and curiosity-driven exploration, using frontier LLMs as the cognitive engine under human-in-the-loop validation. Its published Active Directory walkthrough~\cite{hackersec2026yagaad} describes an agent that progressively builds an understanding of the estate and \emph{maps the path to the Domain Controller} before an attacker does, and the vendor's own benchmark reports Yaga outperforming standalone LLMs on the same tasks~\cite{hackersec2026benchmark}. Horizon3.ai's \textbf{NodeZero} is the other reference point: an independent academic evaluation framework~\cite{prinos2026nodezero} instantiates NodeZero as the autonomous pentester against Microsoft Defender XDR inside a GOAD lab---the same deliberately-vulnerable AD estate we model in Section~\ref{sec:case}---treating its internal-network exploitation as a black-box baseline mature enough to stress a defensive agent. We read both products as strong evidence that robust, context-retaining chain exploitation over an internal network is achievable today: Yaga and NodeZero sustain state across a long engagement because their vendors built dedicated harnesses and infrastructure around the model. Our contribution is orthogonal and complementary: ChainGraph shows that the same externalized, queryable state---the ingredient that lets these agents keep their footing---can be obtained from commodity parts (a filesystem and Neo4j over MCP) wired to an unmodified agent, with no bespoke harness.

\section{Threat and Scope Model}\label{sec:threat}

ChainGraph is a reasoning aid for an \emph{authorized} operator or an autonomous agent acting under explicit authorization. We state assumptions and trust boundaries to bound the contribution.

\textbf{Actors.} (i) The \emph{agent}: an LLM coding tool (Claude Code or Codex) running a ReAct loop, with shell access to offensive tooling and MCP access to Neo4j. (ii) The \emph{operator}: a human with engagement authorization who approves gated actions. (iii) The \emph{target}: the in-scope environment, here a GOAD-modeled AD estate.

\textbf{Trust boundaries.} The agent's model weights and the harness (shell, hooks, MCP client) are trusted. Tool outputs and graph contents are \emph{untrusted data}: they are generated by or derived from the target and must never be executed as instructions---the indirect-prompt-injection boundary~\cite{greshake2023indirect}. The Neo4j instance is a local, loopback-bound reasoning store, not a target asset.

\textbf{In scope.} Externalizing engagement state into a graph; computing paths, severity, and chokepoints; wiring the graph to an unmodified agent; capturing tamper-evident evidence. \textbf{Out of scope.} Automatic, unsupervised exploitation; novel exploit development; evasion of defenses for unauthorized access; and the reliability of a specific model at \emph{populating} the graph from raw output (Section~\ref{sec:future}).

\textbf{Objective.} Given an entry position and a declared \texttt{Objective} node, surface every evidence-backed path to the crown jewel, rank findings by chain-aware severity, and identify the cuts---so a human can prioritize remediation or an authorized agent can plan the next action.

\section{Architecture: Dual Memory}\label{sec:arch}

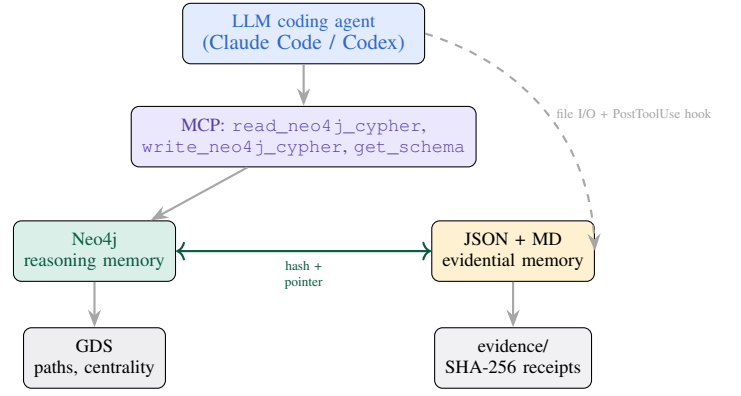
\begin{figure}[t]
\centering
\begin{tikzpicture}[
  font=\scriptsize,
  box/.style={draw, rounded corners=3pt, align=center, minimum height=0.8cm, inner sep=4pt},
  agent/.style={box, fill=graphblue!15, text=graphblue!80!black, minimum width=3.2cm},
  doc/.style={box, fill=nodefind!18, text=black, minimum width=1.5cm},
  graph/.style={box, fill=graphgreen!15, text=graphgreen!60!black, minimum width=1.5cm},
  mcp/.style={box, fill=graphpurple!15, text=graphpurple!70!black, minimum width=3.2cm},
  a/.style={-{Stealth}, thick, gray!70},
]
  \node[agent] (agent) {LLM coding agent\\\footnotesize(Claude Code / Codex)};
  \node[mcp, below=0.55cm of agent] (mcp) {MCP: \texttt{read\_neo4j\_cypher},\\\texttt{write\_neo4j\_cypher}, \texttt{get\_schema}};
  \node[graph, below left=0.7cm and -0.6cm of mcp] (neo) {Neo4j\\\scriptsize reasoning memory};
  \node[doc, below right=0.7cm and -0.6cm of mcp] (fs) {JSON + MD\\\scriptsize evidential memory};
  \node[box, fill=infocolor!15, below=0.6cm of neo, minimum width=1.5cm] (gds) {GDS\\\scriptsize paths, centrality};
  \node[box, fill=infocolor!12, below=0.6cm of fs, minimum width=1.5cm] (ev) {evidence/\\\scriptsize SHA-256 receipts};

  \draw[a] (agent) -- (mcp);
  \draw[a] (mcp) -- (neo);
  \draw[a,dashed] (agent.east) to[bend left=35] node[right,font=\tiny]{file I/O + PostToolUse hook} (fs.east);
  \draw[a] (neo) -- (gds);
  \draw[a] (fs) -- (ev);
  \draw[a, <->, graphgreen!60!black] (neo) -- node[below,font=\tiny,align=center]{hash +\\pointer} (fs);
\end{tikzpicture}
\caption{ChainGraph dual-memory architecture. The agent reaches the reasoning graph through MCP tools and the evidential store through ordinary file I/O; a PostToolUse hook captures and hashes evidence. Graph nodes reference evidence only by hash and path, never by content.}
\label{fig:arch}
\end{figure}

ChainGraph separates two kinds of memory that LLM agents usually cram into one context window (Fig.~\ref{fig:arch}).

\textbf{Documentary / evidential memory} --- the ground truth. Each finding is a JSON document; each narrative artifact (notes, the eventual report) is Markdown. Every observation that justifies a claim is stored as an \emph{evidence receipt}: the raw tool output written to disk, hashed with SHA-256, and referenced by that hash. This layer is append-mostly, diff-friendly under version control, and is the artifact a human auditor reads. It answers \emph{``what did we observe, and can we prove it?''}

\textbf{Relational / reasoning memory} --- the graph. Neo4j holds the same entities as typed nodes, but its value is the \emph{edges}: the pre/post-condition relationships that turn a pile of findings into an attack graph. This layer is queried, recomputed, and traversed. It answers \emph{``what does this enable, where does it lead, and what is the cheapest way to cut it?''}

The two are kept consistent by construction: the graph never stores raw evidence, only a node carrying the evidence's SHA-256 and a path pointer back into the documentary layer (Section~\ref{sec:schema}). An agent that claims a finding can always be challenged to produce the receipt; a receipt that is edited breaks its hash and its graph linkage. This mirrors the evidence-grounding and tamper-evident-accountability properties argued for autonomous harnesses in prior work~\cite{barbosa2026neurosploit}, but obtains them from two commodity stores (a filesystem and Neo4j) rather than a custom engine.

The split also matches how the two stores fail. A filesystem of hashed receipts is durable, auditable, and trivially version-controlled: it is the record that outlives the engagement and the artifact a court or a client reads. A graph is ephemeral working memory: it can be dropped and rebuilt from the receipts at any time (Section~\ref{sec:integration}), so corruption or a bad schema migration costs nothing. Putting evidence in the graph would invert both properties---bloating the store and making the audit record mutable through a query. Keeping them apart means the thing that must be trustworthy (evidence) and the thing that must be fast and disposable (the index) each get the substrate they want.

\paragraph{Why two stores, not one}
A graph database is a poor blob store (large text in properties bloats the page cache) and a filesystem is a poor path finder. More importantly, the split enforces a discipline: the \emph{reasoning} the agent does is always expressed as a query whose inputs are typed and whose output is small, while the \emph{evidence} that grounds a claim is always a hashed artifact a human can open. The separation is what lets us later (Section~\ref{sec:eval}) hand the model a compact query result instead of the whole transcript.

\section{The ChainGraph Schema}\label{sec:schema}

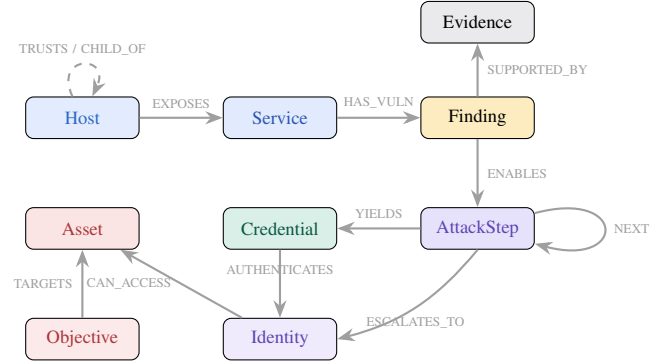
\begin{figure}[t]
\centering
\begin{tikzpicture}[
  font=\scriptsize,
  n/.style={draw, rounded corners=3pt, minimum width=1.5cm, minimum height=0.55cm, align=center, inner sep=2pt},
  host/.style={n, fill=graphblue!15, text=graphblue!80!black},
  svc/.style={n, fill=lowcolor!15, text=lowcolor!70!black},
  find/.style={n, fill=nodefind!25, text=black},
  stepn/.style={n, fill=nodestep!18, text=nodestep!75!black},
  cred/.style={n, fill=nodecred!15, text=nodecred!60!black},
  asset/.style={n, fill=nodeasset!15, text=nodeasset!75!black},
  ev/.style={n, fill=infocolor!18, text=black},
  r/.style={-{Stealth}, thick, gray!75, font=\tiny},
]
  \node[host] (host) {Host};
  \node[svc, right=1.1cm of host] (svc) {Service};
  \node[find, right=1.1cm of svc] (find) {Finding};
  \node[stepn, below=0.9cm of find] (step) {AttackStep};
  \node[cred, left=1.1cm of step] (cred) {Credential};
  \node[asset, left=1.1cm of cred] (asset) {Asset};
  \node[n, fill=graphpurple!12, text=graphpurple!75!black, below=0.9cm of cred] (ident) {Identity};
  \node[ev, above=0.7cm of find] (ev) {Evidence};
  \node[n, fill=critcolor!12, text=critcolor!75!black, below=0.9cm of asset] (obj) {Objective};

  \draw[r] (host) -- node[above]{EXPOSES} (svc);
  \draw[r] (svc) -- node[above]{HAS\_VULN} (find);
  \draw[r] (find) -- node[right]{ENABLES} (step);
  \draw[r] (find) -- node[right]{SUPPORTED\_BY} (ev);
  \draw[r] (step) -- node[above]{YIELDS} (cred);
  \draw[r] (cred) -- node[above]{AUTHENTICATES} (ident);
  \draw[r] (step.south) to[bend left=18] node[below,font=\tiny]{ESCALATES\_TO} (ident.east);
  \draw[r] (ident) -- node[left]{CAN\_ACCESS} (asset);
  \draw[r] (obj) -- node[left]{TARGETS} (asset);
  \draw[r] (step) to[loop right] node[right]{NEXT} (step);
  \draw[r, dashed] (host) to[out=115,in=65,looseness=6] node[above,font=\tiny]{TRUSTS / CHILD\_OF} (host);
\end{tikzpicture}
\caption{ChainGraph property-graph metamodel. Topology edges (\texttt{EXPOSES}, \texttt{TRUSTS}) are discovered; vulnerability edges (\texttt{HAS\_VULN}) attach findings; the attack lattice (\texttt{ENABLES}, \texttt{REQUIRES}, \texttt{YIELDS}, \texttt{ESCALATES\_TO}, \texttt{PIVOTS\_TO}, \texttt{NEXT}) is the reasoning substrate.}
\label{fig:schema}
\end{figure}

The property-graph metamodel is shown in Fig.~\ref{fig:schema}. Nodes carry a single primary label and an \texttt{id}; edges are typed and may carry properties (e.g., a \texttt{TRUSTS} edge records \texttt{sid\_filtering}).

\subsection{Node Types}
\textbf{Finding} is the central record, deliberately richer than a scanner row:
\begin{itemize}[leftmargin=*,nosep]
  \item \texttt{id}, \texttt{category}, \texttt{name}, MITRE technique~\cite{strom2018attack};
  \item \texttt{local\_severity} and \texttt{cvss}~\cite{first2023cvss4} --- risk of the vulnerability \emph{in isolation};
  \item \texttt{confidence} $\in[0,1]$ --- the agent's calibrated belief the finding is real;
  \item \texttt{status} $\in\{$\texttt{confirmed}, \texttt{needs\_review}, \texttt{rejected}$\}$ --- the adjudication gate that keeps unproven claims out of chain computations.
\end{itemize}
\textbf{AttackStep} encodes an atomic action with \texttt{pre}/\texttt{post} conditions, \texttt{technique}, and a qualitative \texttt{risk} (used by the severity model). \textbf{Credential}, \textbf{Identity}, \textbf{Asset}, \textbf{Host}, \textbf{Service} model the environment; \textbf{Objective} marks the engagement goal; \textbf{Evidence} carries only \texttt{tool}, \texttt{sha256}, \texttt{path}, and \texttt{captured\_by}.

\subsection{Edge Types}
The edge vocabulary is the contribution. \texttt{HAS\_VULN} attaches a finding to a service. \texttt{ENABLES} says a finding unlocks an attack step. \texttt{REQUIRES} is the step's precondition (a credential or identity it consumes); \texttt{YIELDS} is its postcondition (a credential it produces). \texttt{AUTHENTICATES} links a credential to the identity it grants. \texttt{ESCALATES\_TO} and \texttt{PIVOTS\_TO} mark privilege and host transitions. \texttt{CAN\_ACCESS} connects an identity to an asset. \texttt{NEXT} records the observed ordering of a concrete kill chain. \texttt{SUPPORTED\_BY} is the provenance edge to evidence. This vocabulary is intentionally small; it is enough to express ``this enables that, which requires this credential, which this step yields, which escalates to that identity, which can access the crown jewel.''

\subsection{Standards on the Edges}
Two external standards ride on the schema rather than being reinvented. Every \texttt{Finding} and \texttt{AttackStep} carries a MITRE ATT\&CK technique ID~\cite{strom2018attack} (e.g.\ \texttt{T1649} for the AD~CS abuse), which makes the attack lattice a concrete instance of the ATT\&CK model and the join key to any technique knowledge graph (Section~\ref{sec:views}). Severity uses CVSS~\cite{first2023cvss4}: the per-finding \texttt{cvss} is the standard base score, and our chain-severity composite (Section~\ref{sec:severity}) is explicitly \emph{not} a CVSS score---it is a path metric layered above it, so the two never get confused. Keeping both as node properties means an agent can filter (``all \texttt{T1649} findings'') or sort (``by CVSS'') with ordinary Cypher, and a report generator can emit standards-compliant references directly from the graph.

A concrete write illustrates how little the agent must say to record a finding and wire it into the lattice (Listing~\ref{lst:write}); this is the single statement step 3--4 of the loop (Table~\ref{tab:loop}) issues through \texttt{write\_neo4j\_cypher}.

\begin{lstlisting}[language=Cypher,caption={The agent's write for one finding and its edges.},label={lst:write}]
MERGE (f:Finding {id:'F-004'})
  SET f += {category:'adcs', technique:'T1649',
            local_severity:'High', cvss:8.1,
            confidence:0.93, status:'confirmed'}
WITH f
MATCH (s:Service {id:'S-DC02-ADCS'})
MATCH (ev:Evidence {id:'EV-004'})
MERGE (s)-[:HAS_VULN]->(f)
MERGE (f)-[:SUPPORTED_BY]->(ev)
WITH f MATCH (k:AttackStep {id:'K-03'})
MERGE (f)-[:ENABLES]->(k);
\end{lstlisting}

\subsection{Constraints and Indexes}
Listing~\ref{lst:schema} shows the schema we applied to Neo4j~5. Uniqueness constraints on \texttt{id} double as lookup indexes and guarantee idempotent \texttt{MERGE} loads; secondary indexes on \texttt{status} and \texttt{category} keep the adjudication filters cheap.

\begin{lstlisting}[language=Cypher,caption={Schema: uniqueness constraints and secondary indexes (excerpt).},label={lst:schema}]
CREATE CONSTRAINT finding_id IF NOT EXISTS
  FOR (n:Finding) REQUIRE n.id IS UNIQUE;
CREATE CONSTRAINT step_id IF NOT EXISTS
  FOR (n:AttackStep) REQUIRE n.id IS UNIQUE;
CREATE CONSTRAINT evidence_id IF NOT EXISTS
  FOR (n:Evidence) REQUIRE n.id IS UNIQUE;
CREATE INDEX finding_status IF NOT EXISTS
  FOR (n:Finding) ON (n.status);
CREATE INDEX evidence_sha IF NOT EXISTS
  FOR (n:Evidence) ON (n.sha256);
\end{lstlisting}

\subsection{A Finding as Data}
Listing~\ref{lst:finding} shows one finding document from the evidential layer. Note what is \emph{not} present: no raw hash, no credential, no tool output---only a pointer to the evidence node whose receipt carries the proof. The \texttt{confidence} and \texttt{status} fields are the calibration hooks; the \texttt{technique} field is the join key to the knowledge overlay (Section~\ref{sec:views}).

\begin{lstlisting}[language=Json,caption={A single Finding document (evidential layer).},label={lst:finding}]
{ "label": "Finding", "id": "F-004",
  "category": "adcs", "technique": "T1649",
  "name": "AD CS template allows enrollee-supplied SAN (ESC1)",
  "local_severity": "High", "cvss": 8.1,
  "confidence": 0.93, "status": "confirmed",
  "evidence": "EV-004" }
\end{lstlisting}

\subsection{Loading}
Findings and relationships are authored as JSON (\texttt{nodes.json}, \texttt{edges.json}) in the evidential layer and loaded with APOC's generic merge procedures (Listing~\ref{lst:load}), so the same data file reconstructs the graph deterministically---important for reproducibility and for rebuilding a scratch reasoning store from version-controlled truth.

\begin{lstlisting}[language=Cypher,caption={Idempotent load via APOC; label and type come from the data.},label={lst:load}]
CALL apoc.load.json("file:///nodes.json") YIELD value AS n
CALL apoc.merge.node([n.label], {id: n.id},
       apoc.map.removeKey(n,"label"), {}) YIELD node
RETURN count(node);

CALL apoc.load.json("file:///edges.json") YIELD value AS r
MATCH (a {id:r.from}) MATCH (b {id:r.to})
CALL apoc.merge.relationship(a, r.type, {},
       coalesce(r.props,{}), b, {}) YIELD rel
RETURN count(rel);
\end{lstlisting}

\section{The Reasoning Layer}\label{sec:reasoning}

The point of the graph is to answer the questions an operator actually asks. We encode five as parameterized, reusable queries. Each is read-only and returns a small result---the property that later yields the token savings.

\subsection{Q1 --- ``What does this vulnerability enable next?''}
From a confirmed finding, follow \texttt{ENABLES} into the step lattice and report the immediate successors and what they yield (Listing~\ref{lst:q1}).

\begin{lstlisting}[language=Cypher,caption={Q1: next-step inference.},label={lst:q1}]
MATCH (f:Finding {id:$finding})-[:ENABLES]->(s:AttackStep)
OPTIONAL MATCH (s)-[:NEXT]->(next:AttackStep)
OPTIONAL MATCH (s)-[:YIELDS|ESCALATES_TO|PIVOTS_TO]->(g)
RETURN f.id AS finding, s.id AS step, s.name AS action,
       next.id AS next_step, collect(DISTINCT g.id) AS gains;
\end{lstlisting}

\subsection{Q2 --- ``What path runs from a low exposure to domain/root?''}
This is the question lists cannot answer. A \texttt{shortestPath} over the enablement/escalation lattice from an entry finding to the objective's crown-jewel asset returns the chain explicitly (Listing~\ref{lst:q2}).

\begin{lstlisting}[language=Cypher,caption={Q2: entry-to-crown-jewel pathing.},label={lst:q2}]
MATCH (entry:Finding {id:$entry})-[:ENABLES]->(s0:AttackStep)
MATCH (obj:Objective)-[:TARGETS]->(crown:Asset)
MATCH p = shortestPath( (s0)
  -[:NEXT|YIELDS|AUTHENTICATES|REQUIRES
    |ESCALATES_TO|PIVOTS_TO|CAN_ACCESS|EXPOSES*1..20]->(crown) )
RETURN entry.id AS entry_finding, crown.name AS crown_jewel,
       length(p) AS hops,
       [n IN nodes(p) | coalesce(n.name,n.id)] AS path
ORDER BY hops ASC LIMIT 3;
\end{lstlisting}

\subsection{Q3 --- ``Which isolated findings turn critical combined?''}\label{sec:severity}
See Section~\ref{sec:case} for output; the model is defined next.

\subsection{Q4 --- ``What evidence sustains this path?''}
Every confirmed finding on the chain, joined to its receipt, so the path is defensible (Listing~\ref{lst:q4}).

\begin{lstlisting}[language=Cypher,caption={Q4: evidence provenance for a path.},label={lst:q4}]
MATCH (f:Finding)-[:ENABLES]->(:AttackStep)
MATCH (f)-[:SUPPORTED_BY]->(ev:Evidence)
WHERE f.status = 'confirmed'
RETURN DISTINCT f.id AS finding, f.name AS claim,
       f.confidence AS conf, ev.tool AS tool,
       left(ev.sha256,12) AS receipt_sha, ev.path AS artifact
ORDER BY finding;
\end{lstlisting}

\subsection{Q5 --- ``Which asset/step is the best chain break-point?''}
We project the attack lattice into GDS and run betweenness centrality~\cite{neo4jgds}: the node lying on the most shortest attack paths is the highest-value remediation (Listing~\ref{lst:q5}).

\begin{lstlisting}[language=Cypher,caption={Q5: GDS projection + betweenness for chokepoints.},label={lst:q5}]
CALL gds.graph.project('attackLattice',
  ['Finding','AttackStep','Credential','Identity',
   'Asset','Host','Service','Evidence','Objective'],
  { ENABLES:{orientation:'NATURAL'}, NEXT:{orientation:'NATURAL'},
    YIELDS:{orientation:'NATURAL'}, REQUIRES:{orientation:'NATURAL'},
    AUTHENTICATES:{orientation:'NATURAL'},
    ESCALATES_TO:{orientation:'NATURAL'},
    PIVOTS_TO:{orientation:'NATURAL'},
    CAN_ACCESS:{orientation:'NATURAL'},
    EXPOSES:{orientation:'NATURAL'} });

CALL gds.betweenness.stream('attackLattice') YIELD nodeId, score
WITH gds.util.asNode(nodeId) AS n, score WHERE score > 0
RETURN labels(n)[0] AS type, coalesce(n.name,n.id) AS node,
       round(score,2) AS betweenness
ORDER BY betweenness DESC LIMIT 8;
\end{lstlisting}

\subsection{The Chain-Severity Model}
The central calibration claim is that \emph{severity is a property of paths, not of findings}. We keep two scores. \emph{Local severity} is the CVSS~\cite{first2023cvss4} of the isolated vulnerability. \emph{Chain severity} is computed over the shortest path the finding opens toward a crown jewel. For a confirmed finding $f$ whose cheapest path $p(f)$ reaches crown jewel $c$, we define a composite
\begin{equation}\label{eq:chain}
\mathrm{chain}(f) = \underbrace{\mathrm{cvss}(f)}_{\text{local}}
  + \underbrace{\beta\cdot\mathbb{1}[c\ \text{crown}]}_{\text{reach bonus}}
  + \underbrace{\sum_{s\in p(f)} w(\mathrm{risk}(s))}_{\text{escalation}}
\end{equation}
with, in our instantiation, $\beta=2.0$ and step weights $w(\texttt{high})=0.6$, $w(\texttt{medium})=0.3$, $w(\texttt{low})=0$. The composite is bucketed: $\ge 9.0\Rightarrow$ Critical, $\ge 7.0\Rightarrow$ High, $\ge 4.0\Rightarrow$ Medium. The exact constants are a policy choice, not a universal law---the contribution is the \emph{form}: a Medium finding that opens a path to the forest root is a Critical chain. Equation~\eqref{eq:chain} is expressed directly in the Q3 Cypher (Listing~\ref{lst:q3}).

Three design choices in the model deserve note. First, only \texttt{confirmed} findings contribute, so an unproven claim can never inflate a chain---calibration is enforced by the \texttt{status} gate before severity is ever computed. Second, the score is computed over the \emph{shortest} path to the crown jewel, i.e.\ the attacker's cheapest option, which is the conservative choice for a defender sizing worst-case blast radius; a variant could score over the stealthiest path (Q6) instead when the threat model is a quiet adversary. Third, the reach bonus $\beta$ is deliberately large relative to per-finding CVSS so that \emph{reaching a crown jewel dominates} the ranking---the whole point is to stop ranking by vulnerability class. This is the graph-native analogue of the classical path-based metrics of Wang et~al.~\cite{wang2008probabilistic} and Homer et~al.~\cite{homer2013aggregating}, computed on demand over an incrementally built graph rather than a pre-enumerated state space, and surfaced to an agent as a single number it did not have to derive.

\begin{lstlisting}[language=Cypher,caption={Q3: chain-severity promotion (core expression).},label={lst:q3}]
MATCH (obj:Objective)-[:TARGETS]->(crown:Asset)
MATCH (f:Finding)-[:ENABLES]->(s0:AttackStep)
WHERE f.status = 'confirmed'
MATCH p = shortestPath( (s0)
  -[:NEXT|YIELDS|AUTHENTICATES|REQUIRES
    |ESCALATES_TO|PIVOTS_TO|CAN_ACCESS|EXPOSES*1..20]->(crown) )
WITH f, crown, length(p) AS hops,
     [n IN nodes(p) WHERE n:AttackStep | n.risk] AS risks
WITH f, crown, hops,
     f.cvss
     + (CASE WHEN crown.crown_jewel THEN 2.0 ELSE 0 END)
     + 0.6*size([r IN risks WHERE r='high'])
     + 0.3*size([r IN risks WHERE r='medium']) AS raw
RETURN f.id AS finding, f.local_severity AS local,
   CASE WHEN raw>=9 THEN 'Critical' WHEN raw>=7 THEN 'High'
        WHEN raw>=4 THEN 'Medium' ELSE 'Low' END AS chain_sev,
   round(raw,1) AS score, hops, crown.name AS reaches
ORDER BY score DESC;
\end{lstlisting}

\subsection{Q6 --- ``What is the stealthiest path, not just the shortest?''}
Shortest-hop is the wrong objective when detection matters. By weighting \texttt{NEXT} edges with a per-step \texttt{detect} cost (\texttt{low}$=1$, \texttt{medium}$=3$, \texttt{high}$=6$) and running GDS weighted Dijkstra~\cite{neo4jgds}, the graph returns the \emph{least-detectable} chain and its cumulative risk (Listing~\ref{lst:q6}). On our data the source-to-cross-forest path accrues a total detection cost of $16.0$ with the per-hop accumulation $[0,1,4,7,13,16]$---the DCSync step ($+6$) is the loudest, quantifying what a defender already suspects. The same mechanism lets an operator trade speed for stealth as an explicit, auditable number rather than model intuition.

\begin{lstlisting}[language=Cypher,caption={Q6: weighted Dijkstra for the stealthiest chain.},label={lst:q6}]
CALL gds.graph.project('stealth', ['AttackStep'],
  {NEXT:{orientation:'NATURAL', properties:'cost'}});
MATCH (src:AttackStep {id:$src}), (dst:AttackStep {id:$dst})
CALL gds.shortestPath.dijkstra.stream('stealth',
  { sourceNode:src, targetNodes:dst,
    relationshipWeightProperty:'cost' })
YIELD totalCost, nodeIds, costs
RETURN totalCost,
       [n IN nodeIds | gds.util.asNode(n).name] AS path, costs;
\end{lstlisting}

\section{Beyond the Attack Graph: Other Graph Views}\label{sec:views}

The same nodes support several \emph{overlay} graphs, each a projection that answers a different class of question (Table~\ref{tab:views}). This is where a property graph earns its keep over a single-purpose attack-graph tool: the views share storage and evidence but differ only in which edge types are traversed.

\begin{table}[t]
\centering
\caption{Graph views over the shared ChainGraph store. Each is a projection over a subset of edge types.}
\label{tab:views}
\footnotesize
\begin{tabular}{@{}p{1.75cm}p{2.9cm}p{2.3cm}@{}}
\toprule
\textbf{View} & \textbf{Edges traversed} & \textbf{Question answered} \\
\midrule
Attack lattice & ENABLES, REQUIRES, YIELDS, ESCALATES\_TO, PIVOTS\_TO, NEXT & Where can the attacker go? \\
Identity / trust & AUTHENTICATES, CAN\_ACCESS, TRUSTS, CHILD\_OF & Who can reach what, across trusts? \\
Asset exposure & EXPOSES, HAS\_VULN & What surface does each host present? \\
Provenance & SUPPORTED\_BY & What proves each claim? \\
Remediation & betweenness over the lattice & Which cut breaks the most chains? \\
\bottomrule
\end{tabular}
\end{table}

\textbf{Identity/trust view} is the BloodHound~\cite{robbins2016bloodhound} question---six degrees of Domain Admin---restricted to \texttt{AUTHENTICATES}, \texttt{CAN\_ACCESS}, and cross-forest \texttt{TRUSTS}/\texttt{CHILD\_OF} edges. ChainGraph subsumes it: BloodHound's identity graph is one projection of our store, and because our credentials carry provenance, an access edge is backed by the evidence that demonstrated it.

Listing~\ref{lst:identity} is the identity-view query: every identity that can reach a crown jewel, via credentials and cross-forest trusts---``who is effectively enterprise admin?''---answered over the same store that answers the attacker's pathing.

\begin{lstlisting}[language=Cypher,caption={Identity view: principals that can reach a crown jewel.},label={lst:identity}]
MATCH (obj:Objective)-[:TARGETS]->(crown:Asset)
MATCH (i:Identity)
MATCH path = shortestPath(
  (i)-[:CAN_ACCESS|AUTHENTICATES|TRUSTS|CHILD_OF|PIVOTS_TO*1..12]-(crown))
RETURN i.name AS principal, i.tier AS tier, length(path) AS hops
ORDER BY hops ASC;
\end{lstlisting}

\textbf{Provenance view} (traverse only \texttt{SUPPORTED\_BY}) is the auditor's graph: from any claim to its receipts, or inverted, from a tampered receipt to every claim it would invalidate. This is the graph-native form of the tamper-evident accountability property~\cite{barbosa2026neurosploit}.

\textbf{Technique knowledge overlay.} Findings and steps carry MITRE technique IDs~\cite{strom2018attack}; joining them to an external technique knowledge graph~\cite{zhang2025attackgplus,prapty2024ragattackgraph} lets the agent ground a step in documented tradecraft and retrieve detection/mitigation guidance---a security-domain instance of GraphRAG~\cite{edge2024graphrag,pan2024unifying}.

\textbf{Remediation view} inverts the attacker's betweenness ranking (Q5) into a defender's work queue: fix the highest-centrality node first, re-run betweenness, repeat. The graph turns an unordered finding list into a prioritized, path-aware remediation plan.

\section{Integration: No Harness Required}\label{sec:integration}

A central practical claim is that ChainGraph needs no custom agent runtime. The reasoning graph is reachable from an unmodified coding agent through the Neo4j Cypher MCP server~\cite{neo4jmcp}, which exposes three tools: \texttt{get\_neo4j\_schema}, \texttt{read\_neo4j\_cypher}, and \texttt{write\_neo4j\_cypher}.

\subsection{Step 1 --- Bring up Neo4j}
Listing~\ref{lst:compose} is the full Docker Compose service. We bind the ports to loopback only, enable APOC and GDS via the documented \texttt{NEO4J\_PLUGINS} key, and set heap/page-cache through the Neo4j~5 environment-variable convention (dots become \texttt{\_}, underscores in a setting name double)~\cite{neo4jgds}.

\begin{lstlisting}[language=Json,caption={docker-compose.yml for the reasoning store (loopback-bound).},label={lst:compose}]
services:
  neo4j:
    image: neo4j:5-community
    ports: ["127.0.0.1:7474:7474","127.0.0.1:7687:7687"]
    environment:
      NEO4J_AUTH: "neo4j/${NEO4J_PASSWORD:?set it}"
      NEO4J_PLUGINS: '["apoc","graph-data-science"]'
      NEO4J_apoc_import_file_enabled: "true"
      NEO4J_server_directories_import: "/import"
      NEO4J_server_memory_heap_max__size: "2G"
      NEO4J_server_memory_pagecache_size: "1G"
\end{lstlisting}

\subsection{Step 2 --- Register the MCP server}
For Claude Code, one command registers the server at project scope, writing \texttt{.mcp.json}~\cite{claudecodedocs} (Listing~\ref{lst:claudeadd}). The \texttt{NEO4J\_READ\_ONLY} variable and tool-level deny rules give us defense in depth because Neo4j Community Edition has no role-based access control---read-only database users are an Enterprise-only feature~\cite{neo4jgds}, so we enforce the read/write boundary at the protocol layer instead.

\begin{lstlisting}[language=Shell,caption={Registering the Neo4j MCP server with Claude Code.},label={lst:claudeadd}]
claude mcp add neo4j --scope project \
  -e NEO4J_URI=bolt://127.0.0.1:7687 \
  -e NEO4J_USERNAME=neo4j \
  -e NEO4J_PASSWORD="$NEO4J_PASSWORD" \
  -e NEO4J_DATABASE=neo4j \
  -- uvx mcp-neo4j-cypher@0.3 --transport stdio
\end{lstlisting}

The resulting tools are named \texttt{mcp\_\_neo4j\_\_read\_neo4j\_cypher} and \texttt{mcp\_\_neo4j\_\_write\_neo4j\_cypher}. We allow the read tool broadly and gate the write tool behind explicit approval with a permission rule (Listing~\ref{lst:perms}); because MCP deny rules cannot match on parameters~\cite{claudecodedocs}, we deny the write tool wholesale and let the agent request it per use, which is exactly the authorization gate a state-changing operation deserves.

\begin{lstlisting}[language=Json,caption={Permission policy: read freely, gate writes (.claude/settings.json).},label={lst:perms}]
{ "permissions": {
    "allow": ["mcp__neo4j__read_neo4j_cypher",
              "mcp__neo4j__get_neo4j_schema"],
    "ask":   ["mcp__neo4j__write_neo4j_cypher"] } }
\end{lstlisting}

\subsection{Step 3 --- Capture evidence automatically}
A PostToolUse hook turns every tool run into an evidence receipt without the model's cooperation. The hook receives JSON on stdin with \texttt{tool\_name}, \texttt{tool\_input}, and \texttt{tool\_response}~\cite{claudecodedocs}; it writes the output to \texttt{evidence/}, hashes it, and (optionally) \texttt{MERGE}s an \texttt{Evidence} node (Listing~\ref{lst:hook}). Because the receipt is produced by the harness, not narrated by the LLM, a finding's provenance cannot be hallucinated.

\begin{lstlisting}[language=Json,caption={PostToolUse hook wiring (settings.json).},label={lst:hook}]
{ "hooks": { "PostToolUse": [ {
   "matcher": "Bash",
   "hooks": [ { "type": "command",
     "command": ".claude/hooks/capture_evidence.sh" } ] } ] } }
\end{lstlisting}

The capture script itself is short (Listing~\ref{lst:hookscript}): read the hook JSON from stdin, extract the command and its output, write the receipt, hash it, and print nothing (a silent hook does not perturb the agent). Writing the \texttt{Evidence} node is optional here and can be deferred to the agent's next \texttt{write\_neo4j\_cypher} so that all graph writes flow through one gated path.

\begin{lstlisting}[language=Shell,caption={capture\_evidence.sh: a silent PostToolUse evidence recorder.},label={lst:hookscript}]
#!/usr/bin/env bash
# stdin: JSON with tool_name, tool_input, tool_response
in=$(cat); id="EV-$(date +%s%N)"
dir="evidence/$id"; mkdir -p "$dir"
jq -r '.tool_input.command // ""'  <<<"$in" > "$dir/cmd.txt"
jq -r '.tool_response.stdout // ""' <<<"$in" > "$dir/stdout.txt"
sha256sum "$dir/stdout.txt" | awk '{print $1}' > "$dir/sha256"
exit 0   # silent: no stdout -> no perturbation of the agent
\end{lstlisting}

\subsection{Step 4 --- Codex uses the same graph}
OpenAI Codex~\cite{codexdocs} reaches the identical Neo4j MCP server through \texttt{\textasciitilde/.codex/config.toml} (Listing~\ref{lst:codex}); project guidance lives in \texttt{AGENTS.md} rather than \texttt{CLAUDE.md}. The graph, the schema, and the five queries are agent-agnostic: switching the driver model changes nothing in the reasoning layer.

\begin{lstlisting}[language=Toml,caption={Codex MCP configuration (config.toml).},label={lst:codex}]
[mcp_servers.neo4j]
command = "uvx"
args = ["mcp-neo4j-cypher@0.3", "--transport", "stdio"]
env = { NEO4J_URI = "bolt://127.0.0.1:7687",
        NEO4J_USERNAME = "neo4j",
        NEO4J_PASSWORD = "...",
        NEO4J_DATABASE = "neo4j",
        NEO4J_READ_ONLY = "true" }
\end{lstlisting}

\begin{figure*}[t]
\centering
\resizebox{0.96\textwidth}{!}{%
\begin{tikzpicture}[
  every node/.style={font=\scriptsize},
  gnode/.style={circle, draw=black!55, line width=0.3pt, minimum size=4.2mm, inner sep=0pt},
  host/.style={gnode, fill=blue!55},
  find/.style={gnode, fill=yellow!80!orange},
  astep/.style={gnode, fill=violet!60},
  cred/.style={gnode, fill=green!60!black!70},
  idn/.style={gnode, fill=purple!55},
  asset/.style={gnode, fill=red!65},
  obj/.style={circle, draw=black, line width=0.6pt, fill=orange, minimum size=8mm, inner sep=0pt, font=\scriptsize\bfseries},
  gedge/.style={-{Latex[length=1.3mm]}, draw=black!35, line width=0.25pt},
  collabel/.style={font=\scriptsize\bfseries, text=black!70}]

  \def\xh{0} \def\xf{2.6} \def\xs{5.2} \def\xc{7.8} \def\xi{10.4} \def\xa{13.0} \def\xo{15.6}

  \foreach \i/\y in {1/0.6,2/2.0,3/3.4,4/4.8} { \node[host] (h\i) at (\xh,\y) {}; }
  \foreach \i/\y in {1/0.2,2/1.3,3/2.4,4/3.5,5/4.6,6/5.4} { \node[find] (f\i) at (\xf,\y) {}; }
  \foreach \i/\y in {1/0.8,2/2.1,3/3.4,4/4.7} { \node[astep] (s\i) at (\xs,\y) {}; }
  \foreach \i/\y in {1/1.2,2/2.7,3/4.2} { \node[cred] (c\i) at (\xc,\y) {}; }
  \foreach \i/\y in {1/1.6,2/3.0,3/4.4} { \node[idn] (i\i) at (\xi,\y) {}; }
  \foreach \i/\y in {1/2.0,2/3.6} { \node[asset] (a\i) at (\xa,\y) {}; }
  \node[obj] (O) at (\xo,2.8) {OBJ};

  \draw[gedge] (h1)--(f1); \draw[gedge] (h1)--(f2);
  \draw[gedge] (h2)--(f3); \draw[gedge] (h2)--(f2);
  \draw[gedge] (h3)--(f4); \draw[gedge] (h3)--(f5);
  \draw[gedge] (h4)--(f6); \draw[gedge] (h4)--(f5);
  \draw[gedge] (f1)--(s1); \draw[gedge] (f2)--(s1);
  \draw[gedge] (f3)--(s2); \draw[gedge] (f4)--(s3);
  \draw[gedge] (f5)--(s4); \draw[gedge] (f6)--(s4); \draw[gedge] (f2)--(s2);
  \draw[gedge] (s1)--(c1); \draw[gedge] (s2)--(c2);
  \draw[gedge] (s3)--(c2); \draw[gedge] (s4)--(c3); \draw[gedge] (s2)--(c1);
  \draw[gedge] (c1)--(i1); \draw[gedge] (c2)--(i2);
  \draw[gedge] (c3)--(i3); \draw[gedge] (c2)--(i1);
  \draw[gedge] (i1)--(a1); \draw[gedge] (i2)--(a1);
  \draw[gedge] (i2)--(a2); \draw[gedge] (i3)--(a2);
  \draw[gedge, draw=black!55, line width=0.4pt] (a1)--(O);
  \draw[gedge, draw=black!55, line width=0.4pt] (a2)--(O);
  \draw[gedge] (i1) to[bend left=12] (i2);
  \draw[gedge] (s3) to[bend right=10] (c3);

  \node[collabel] at (\xh,6.1) {Host/Service};
  \node[collabel] at (\xf,6.1) {Finding};
  \node[collabel] at (\xs,6.1) {AttackStep};
  \node[collabel] at (\xc,6.1) {Credential};
  \node[collabel] at (\xi,6.1) {Identity};
  \node[collabel] at (\xa,6.1) {Asset};
  \node[collabel] at (\xo,6.1) {Objective};

  \node[font=\tiny, text=black!55] at (1.3,5.0) {EXPOSES};
  \node[font=\tiny, text=black!55] at (3.9,4.3) {ENABLES};
  \node[font=\tiny, text=black!55] at (6.5,3.9) {YIELDS};
  \node[font=\tiny, text=black!55] at (9.1,4.1) {AUTHENTICATES};
  \node[font=\tiny, text=black!55] at (11.7,3.3) {ESCALATES\_TO};
  \node[font=\tiny, text=black!55] at (14.3,3.6) {CAN\_ACCESS};
\end{tikzpicture}}
\caption{Schematic of the ChainGraph the agent reasons over (structure of the engagement-scale store; node counts in Section~\ref{sec:case}). Typed nodes---hosts/services (blue), findings (yellow), attack steps (violet), credentials (green), identities (purple), assets (red)---connect through directed, pre/post-condition edges and converge on the central \texttt{Objective} (orange). What a flat-context agent would hold as a line-by-line transcript, this agent holds as a graph it queries: a path from any leaf to the center is one \texttt{shortestPath} call, not a re-read.}
\label{fig:livegraph}
\end{figure*}
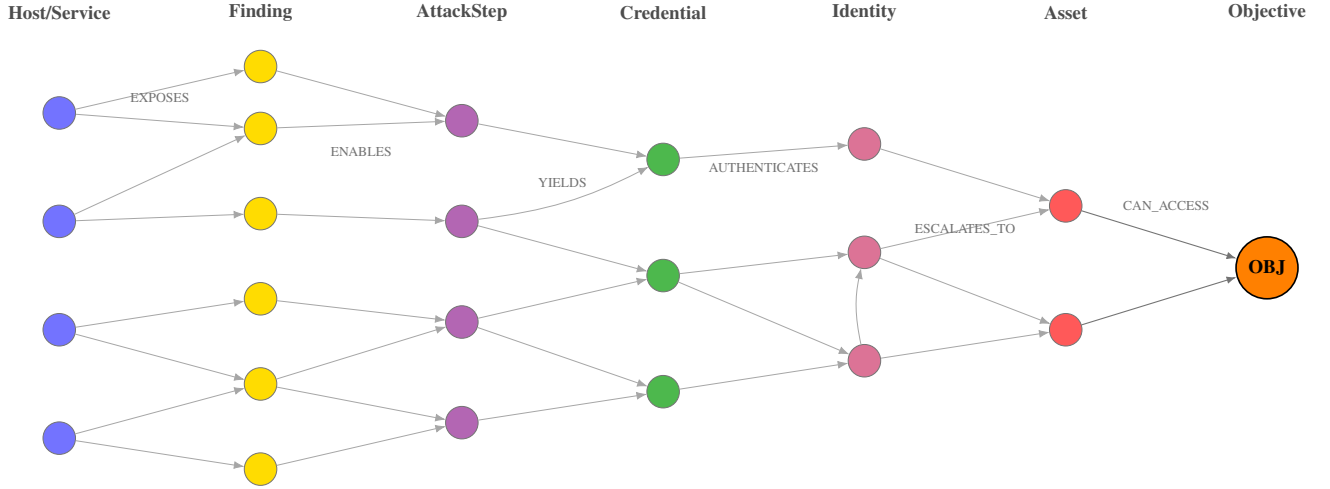

\subsection{Kit Layout and Reproducibility}
The kit is a single directory: \texttt{graph/} (schema and APOC loader), \texttt{queries/} (the six \texttt{.cypher} files), \texttt{engagement/} (the version-controlled \texttt{nodes.json}, \texttt{edges.json}, and \texttt{evidence/} receipts), \texttt{bin/make\_scenario.py} (generates the synthetic engagement deterministically), \texttt{docker-compose.yml}, and \texttt{.claude/} (settings, hook). Reconstructing the entire reasoning graph from scratch is three commands: \texttt{docker compose up -d}; copy the JSON into the import dir; run the schema and loader. Because the graph is a pure function of the JSON, a reviewer rebuilds the exact 48-node, 50-edge store used for every number in Section~\ref{sec:case}. The evidential layer is the source of truth; the graph is a disposable, rebuildable index over it---the same discipline that lets an engagement survive a context compaction.

\subsection{The Engagement Loop}
With the wiring in place, the agent runs a ReAct loop~\cite{yao2023react} in which the graph is consulted at every decision: after a tool runs, the hook records evidence and the agent \texttt{write\_neo4j\_cypher}s the new finding/step/edges; before choosing the next action, the agent \texttt{read\_neo4j\_cypher}s Q1/Q2 to see what the latest finding unlocks and whether a path to the objective now exists; for triage, it runs Q3/Q5. The model authors these queries itself---Text2Cypher-style~\cite{ozsoy2024text2cypher,feng2024cypherbench}---or reuses the saved parameterized ones.

Table~\ref{tab:loop} traces one cycle of the loop over the case-study data: the agent captures evidence, writes a finding and its edges, then reads back the path that the new finding created. The write-then-read discipline is what keeps the model's working set small---it never holds the whole engagement, only the delta it just wrote and the answer it just asked for.

\begin{table}[t]
\centering
\caption{One engagement-loop cycle: MCP tool calls around a single finding. (R)=read tool, (W)=write tool, (H)=hook, (F)=file I/O.}
\label{tab:loop}
\footnotesize
\begin{tabular}{@{}llp{4.1cm}@{}}
\toprule
\textbf{\#} & \textbf{Actor} & \textbf{Action} \\
\midrule
1 & agent+Bash & run \texttt{certipy find -vulnerable} \\
2 & hook (H)   & write output to \texttt{evidence/EV-004/}, SHA-256 it \\
3 & agent (W)  & \texttt{MERGE (:Finding \{id:'F-004',\dots\})}, \texttt{SUPPORTED\_BY} $\rightarrow$ \texttt{EV-004} \\
4 & agent (W)  & \texttt{MERGE} \texttt{ENABLES} edges to steps K-03, K-04 \\
5 & agent (R)  & Q1(F-004): returns next steps + gains \\
6 & agent (R)  & Q2(F-004): path to crown jewel now exists (4 hops) \\
7 & agent (R)  & Q3: F-004 ranks Critical (score 11.3) \\
8 & agent      & decide next action from the 3 small results, not the transcript \\
\bottomrule
\end{tabular}
\end{table}

\section{Case Study: A Multi-Forest AD Chain}\label{sec:case}

\begin{figure}[t]
\centering
\begin{tikzpicture}[
  font=\scriptsize, node distance=0.9cm and 1.4cm,
  dc/.style={draw, rounded corners=3pt, fill=graphblue!15, minimum width=1.8cm, minimum height=0.7cm, align=center, text=graphblue!85!black},
  srv/.style={draw, rounded corners=3pt, fill=nodeasset!15, minimum width=1.7cm, minimum height=0.7cm, align=center, text=nodeasset!80!black},
  forest/.style={draw, dashed, rounded corners=7pt, inner sep=7pt},
  trust/.style={<->, thick, >=Stealth, color=highcolor!80!black},
  child/.style={->, thick, >=Stealth, color=graphblue!70},
]
  \node[dc] (dc01) {DC01\\\scriptsize Winterfell};
  \node[dc, below right=0.7cm and 1.3cm of dc01] (dc02) {DC02\\\scriptsize North};
  \node[srv, below=0.7cm of dc01] (srv03) {SRV03\\\scriptsize Web/Shares};
  \node[srv, below=0.7cm of dc02] (srv02) {SRV02\\\scriptsize MSSQL/IIS};
  \begin{scope}[on background layer]
    \node[forest, fill=graphblue!5, label={[font=\tiny\bfseries,text=graphblue]above:sevenkingdoms.local}] (f1) [fit=(dc01)(dc02)(srv02)(srv03)] {};
  \end{scope}
  \node[dc, right=3.1cm of dc01] (dc03) {DC03\\\scriptsize Meereen};
  \begin{scope}[on background layer]
    \node[forest, fill=nodeasset!5, label={[font=\tiny\bfseries,text=nodeasset]above:essos.local}] (f2) [fit=(dc03)] {};
  \end{scope}
  \draw[child] (dc01) -- node[above,sloped,font=\tiny]{parent} (dc02);
  \draw[trust] (dc01) -- node[above,font=\tiny]{bidir.\ trust, no SID filter} (dc03);
\end{tikzpicture}
\caption{GOAD-modeled topology used in the case study: two forests, three domains, five hosts, with a cross-forest trust that enables the final pivot.}
\label{fig:topo}
\end{figure}
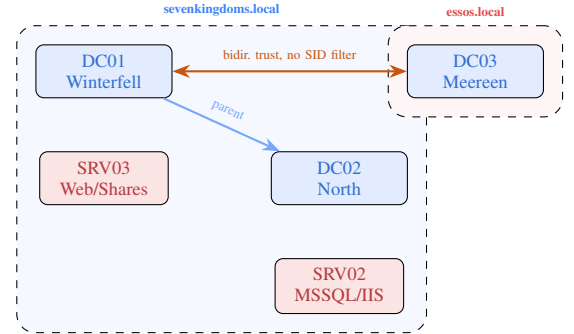

We instantiate ChainGraph on a synthetic engagement modeled on GOAD~\cite{goad}: two forests (\texttt{sevenkingdoms.local}, \texttt{essos.local}), three domains, five hosts, with a cross-forest trust lacking SID filtering (Fig.~\ref{fig:topo}).

The data are abstract descriptors---no exploit code, credentials, or verbatim tool output are stored; \texttt{Evidence} nodes carry only hashes---so the artifact is safe to publish and reproduce. We loaded it into a live Neo4j~5.26.31 instance with GDS~2.13.13 and APOC~5.26.31. The resulting graph has \textbf{48 nodes and 50 edges} (Table~\ref{tab:graphstats}); of ten findings, eight are \texttt{confirmed}, one is \texttt{needs\_review} (a suspected AS-REP-roastable account at confidence 0.40), and one is \texttt{rejected} (an unreproducible web panel at confidence 0.15). Only confirmed findings enter chain computations.

\begin{table}[t]
\centering
\caption{Loaded ChainGraph for the GOAD-modeled engagement.}
\label{tab:graphstats}
\footnotesize
\begin{tabular}{@{}lr lr@{}}
\toprule
\textbf{Node label} & \textbf{n} & \textbf{Edge type} & \textbf{n} \\
\midrule
Finding    & 10 & SUPPORTED\_BY & 10 \\
Evidence   & 10 & HAS\_VULN     & 7  \\
AttackStep & 6  & ENABLES       & 6  \\
Host       & 5  & EXPOSES       & 5  \\
Identity   & 5  & REQUIRES      & 5  \\
Service    & 4  & NEXT          & 5  \\
Credential & 4  & YIELDS        & 3  \\
Asset      & 3  & AUTHENTICATES & 2  \\
Objective  & 1  & others        & 7  \\
\midrule
\textbf{Total nodes} & \textbf{48} & \textbf{Total edges} & \textbf{50} \\
\bottomrule
\end{tabular}
\end{table}

\subsection{Q2 in Action: The Chain}
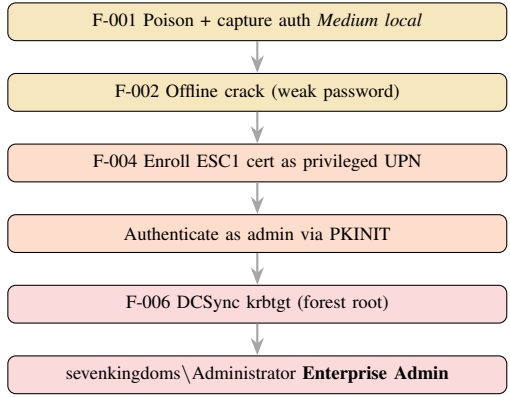
\begin{figure}[t]
\centering
\begin{tikzpicture}[
  font=\scriptsize, node distance=0.42cm,
  s/.style={draw, rounded corners=3pt, fill=nodestep!15, text=nodestep!80!black, align=center, minimum width=6.6cm, minimum height=0.5cm, inner sep=2pt},
  m/.style={fill=medcolor!25, text=black},
  h/.style={fill=highcolor!22, text=black},
  c/.style={fill=critcolor!20, text=black},
  a/.style={-{Stealth}, thick, gray!70},
]
  \node[s,m] (n1) {F-001 Poison + capture auth \hfill \textit{Medium local}};
  \node[s,m, below=of n1] (n2) {F-002 Offline crack (weak password)};
  \node[s,h, below=of n2] (n3) {F-004 Enroll ESC1 cert as privileged UPN};
  \node[s,h, below=of n3] (n4) {Authenticate as admin via PKINIT};
  \node[s,c, below=of n4] (n5) {F-006 DCSync krbtgt (forest root)};
  \node[s,c, below=of n5] (n6) {sevenkingdoms$\backslash$Administrator \hfill \textbf{Enterprise Admin}};
  \foreach \a/\b in {n1/n2,n2/n3,n3/n4,n4/n5,n5/n6} \draw[a] (\a)--(\b);
\end{tikzpicture}
\caption{The Q2 shortest path from a Medium-severity entry finding to forest-root control, as returned by Neo4j. Node color = chain contribution (yellow Medium, orange High, red Critical).}
\label{fig:chain}
\end{figure}

Querying Q2 from the lowest-exposure entry---\texttt{F-001}, LLMNR/NBT-NS poisoning, rated Medium (CVSS 5.9)---returns an eight-node, seven-hop path to the forest root (Fig.~\ref{fig:chain}):
\begin{quote}\footnotesize\itshape
Poison + capture auth $\rightarrow$ Offline crack $\rightarrow$ Enroll ESC1 cert as privileged UPN $\rightarrow$ Authenticate as admin via PKINIT $\rightarrow$ DCSync krbtgt $\rightarrow$ krbtgt hash $\rightarrow$ sevenkingdoms$\backslash$Administrator $\rightarrow$ forest root.
\end{quote}
This is the literal result row returned by Neo4j, not a narrative reconstruction. The chain crosses a protocol-poisoning finding (Responder-class~\cite{responder}), a credential-cracking step, an AD~CS ESC1 abuse~\cite{schroeder2021certified}, a PKINIT authentication, and DCSync---techniques a list-shaped agent would report as five unrelated findings.

\subsection{Q3 in Action: Severity Promotion}
Table~\ref{tab:promo} is the verbatim Q3 output. The headline result: \textbf{three findings rated Medium in isolation---F-001, F-002, F-007---are promoted to Critical chain severity}, because each opens a shortest path that terminates at the forest-root crown jewel. F-007 (cross-forest trust without SID filtering, CVSS 6.8) is the starkest: two hops from a crown jewel, it is a two-line note in a scanner report and a Critical in the graph.

\begin{table}[t]
\centering
\caption{Q3 output: local vs.\ chain severity (verbatim). All confirmed findings reach the forest-root crown jewel; Mediums are promoted.}
\label{tab:promo}
\footnotesize
\begin{tabular}{@{}llccl@{}}
\toprule
\textbf{Finding} & \textbf{Local} & \textbf{Chain} & \textbf{Score} & \textbf{Hops} \\
\midrule
F-006 DCSync          & High   & \textcolor{critcolor}{\textbf{Critical}} & 11.7 & 3 \\
F-004 AD CS ESC1      & High   & \textcolor{critcolor}{\textbf{Critical}} & 11.3 & 5 \\
F-004 (alt.\ path)    & High   & \textcolor{critcolor}{\textbf{Critical}} & 11.0 & 4 \\
F-001 LLMNR poison    & \textcolor{medcolor}{\textbf{Medium}} & \textcolor{critcolor}{\textbf{Critical}} & 9.1 & 7 \\
F-002 Offline crack   & \textcolor{medcolor}{\textbf{Medium}} & \textcolor{critcolor}{\textbf{Critical}} & 9.1 & 6 \\
F-007 Cross-forest    & \textcolor{medcolor}{\textbf{Medium}} & \textcolor{critcolor}{\textbf{Critical}} & 9.1 & 2 \\
\bottomrule
\end{tabular}
\end{table}

\subsection{Q5 in Action: Chokepoints}
GDS projected 48 nodes and the 30 lattice relationships and computed betweenness (Table~\ref{tab:choke}, verbatim). The two highest-centrality nodes are the \emph{Authenticate as admin via PKINIT} step (54.5) and \emph{DCSync krbtgt} (50.0): they lie on the most attack paths and are therefore the best remediation targets---hardening AD~CS enrollment and tiering DCSync rights cuts more chains than fixing any single entry finding. This is the defensive payoff of attacker-side graph thinking: the agent does not just enumerate paths, it ranks the cuts.

\begin{figure}[t]
\centering
\begin{tikzpicture}
\begin{axis}[
  width=\columnwidth, height=4.4cm,
  xbar, bar width=7pt, xmin=0, xmax=62,
  symbolic y coords={DC03, {sk\textbackslash Admin}, Poison, Cross-forest, Crack, ESC1-enroll, DCSync, PKINIT-auth},
  ytick=data, y dir=normal,
  nodes near coords, nodes near coords align={horizontal},
  every node near coord/.append style={font=\tiny},
  xlabel={\scriptsize betweenness centrality (GDS, exact)},
  tick label style={font=\tiny}, xlabel style={font=\scriptsize},
  enlarge y limits=0.08,
]
\addplot[fill=critcolor!75] coordinates {
  (54.5,PKINIT-auth) (50,DCSync) (47.5,ESC1-enroll) (42,Crack)
  (31,Cross-forest) (16,Poison) (12,{sk\textbackslash Admin}) (11,DC03) };
\end{axis}
\end{tikzpicture}
\caption{Q5 betweenness centrality over the attack lattice (real GDS output). The PKINIT-authentication and DCSync steps dominate: cutting them breaks the most chains.}
\label{fig:choke}
\end{figure}
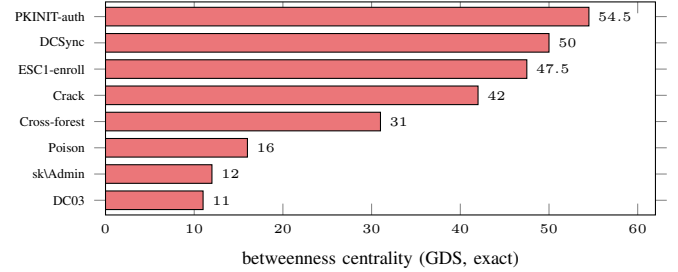

\begin{table}[t]
\centering
\caption{Q5 betweenness centrality (verbatim top rows): the chain break-points.}
\label{tab:choke}
\footnotesize
\begin{tabular}{@{}llr@{}}
\toprule
\textbf{Type} & \textbf{Node} & \textbf{Betweenness} \\
\midrule
AttackStep & Authenticate as admin via PKINIT & 54.5 \\
AttackStep & DCSync krbtgt                    & 50.0 \\
AttackStep & Enroll ESC1 cert as privileged UPN & 47.5 \\
AttackStep & Offline crack                    & 42.0 \\
AttackStep & Cross-forest hop via trust       & 31.0 \\
AttackStep & Poison + capture auth            & 16.0 \\
Identity   & sevenkingdoms\textbackslash Administrator & 12.0 \\
Host       & DC03                             & 11.0 \\
\bottomrule
\end{tabular}
\end{table}

\subsection{Q4 in Action: Provenance}
Q4 returns each confirmed chain finding with its tool and a truncated receipt hash (e.g., F-001 $\leftarrow$ \texttt{responder}, \texttt{bf13d30f1ad2}; F-006 $\leftarrow$ \texttt{secretsdump}, \texttt{ce09c1557e29}). Because the hash is computed by the capture hook over the raw artifact, the path in Fig.~\ref{fig:chain} is defensible end to end: every edge traces to a receipt a human can re-open and re-hash.

\subsection{Three Paths the Graph Separates}
A list-shaped agent reports the ten findings as ten rows. The graph reports them as a small number of \emph{paths}, and distinct paths demand distinct remediations.

\textbf{Path A --- poison to forest root (the promoted Medium).} F-001 (LLMNR) $\rightarrow$ F-002 (crack) $\rightarrow$ F-004 (ESC1) $\rightarrow$ PKINIT $\rightarrow$ F-006 (DCSync). Seven hops, every link evidence-backed, terminating at the \texttt{krbtgt} hash and enterprise-admin identity. The entry finding is Medium in a scanner; the chain is Critical (Table~\ref{tab:promo}). Remediation at the entry (disable LLMNR) is cheap but leaves parallel paths; the graph shows why the chokepoint fix (AD~CS) matters more.

\textbf{Path B --- the short, loud one.} F-004 (ESC1) $\rightarrow$ PKINIT $\rightarrow$ F-006 (DCSync), four hops, chain score 11.0--11.3. This is the path a time-pressured attacker takes and a monitored defender catches first; Q6 assigns it a higher cumulative detection cost at the DCSync step. The graph holds both the fast-and-loud and slow-and-quiet variants simultaneously.

\textbf{Path C --- the cross-forest extension.} Once the forest root is owned, F-007 (cross-forest trust without SID filtering) $\rightarrow$ cross-forest hop $\rightarrow$ \texttt{essos.local}. Two hops from a crown jewel, F-007 is a low-noise note in isolation and the edge that turns a single-forest compromise into a two-forest one. Only a path-aware view promotes it (Table~\ref{tab:promo}).

\section{Efficiency, Memory, and Calibration}\label{sec:eval}

\subsection{Reasoning-Payload Reduction}
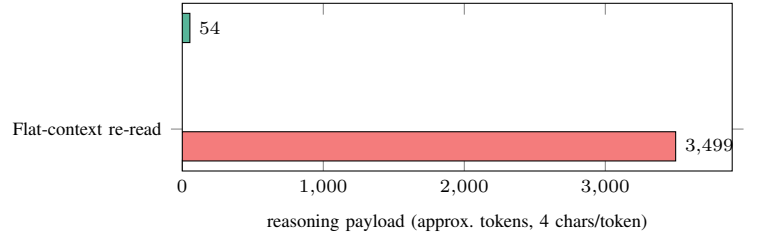
\begin{figure}[t]
\centering
\begin{tikzpicture}
\begin{axis}[
  width=\columnwidth, height=3.8cm,
  xbar, bar width=11pt,
  xmin=0, xmax=3900,
  symbolic y coords={Graph query result, Flat-context re-read},
  ytick=data, y dir=reverse,
  nodes near coords, nodes near coords align={horizontal},
  every node near coord/.append style={font=\scriptsize},
  xlabel={\scriptsize reasoning payload (approx.\ tokens, 4 chars/token)},
  tick label style={font=\scriptsize}, xlabel style={font=\scriptsize},
  enlarge y limits=0.5,
]
\addplot[fill=directred!70] coordinates {(3499,Flat-context re-read)};
\addplot[fill=graphgreen!70] coordinates {(54,Graph query result)};
\end{axis}
\end{tikzpicture}
\caption{Illustrative payload for one path query: handing back a graph result versus re-reading the engagement corpus ($\approx 98\%$ fewer tokens on a character proxy).}
\label{fig:tokens}
\end{figure}

The flat-context agent answers ``is there a path from F-001 to the forest root?'' by re-reading the engagement corpus. Our nodes/edges corpus is $13{,}995$ characters---$\approx 3{,}500$ tokens on a conservative four-characters-per-token proxy---and in a real engagement the transcript (tool output, deliberation) is far larger. ChainGraph answers the same question with one \texttt{shortestPath} query whose result---the eight-node path row---is $\approx 54$ tokens. On the proxy that is a \textbf{$\approx 98\%$ reduction} in the reasoning payload for that decision (Fig.~\ref{fig:tokens}). The mechanism is simple: the graph does the traversal in the database engine and hands back only the answer, so the model never re-attends to the raw corpus~\cite{liu2024lost}. We stress this is a character-based proxy on a single query, offered to illustrate the mechanism, not a tokenizer-exact or multi-run benchmark (Section~\ref{sec:limits}).

\subsection{Why the Gap Widens with Scale}
The flat-context cost grows with the engagement; the graph-query cost grows only with the answer. Table~\ref{tab:scale} projects the per-decision payload for the path question as the engagement accumulates findings, holding the answer (one path row) roughly constant. The flat approach also eventually exceeds the context window and forces lossy compaction---precisely when the lost-in-the-middle degradation~\cite{liu2024lost} is worst---whereas the graph is unaffected because it is never resident in the window.

\begin{table}[t]
\centering
\caption{Projected per-decision reasoning payload for the ``is there a path?'' question as the engagement grows (character-proxy tokens). Graph answer stays $\approx$ one path row.}
\label{tab:scale}
\footnotesize
\begin{tabular}{@{}lrrr@{}}
\toprule
\textbf{Engagement size} & \textbf{Flat ctx.} & \textbf{Graph} & \textbf{Reduction} \\
\midrule
Case study (48 nodes)      & $\approx$3{,}500  & $\approx$54 & 98\% \\
Medium (10$\times$ data)   & $\approx$35{,}000 & $\approx$70 & $>$99\% \\
Large (exceeds window)     & compaction req.\  & $\approx$90 & --- \\
\bottomrule
\end{tabular}
\end{table}

\subsection{Three Compounding Gains}
\textbf{Token/cost.} Per-decision retrieval scales with answer size, not history size, so cost stays flat as the engagement grows---unlike a context window that must carry the whole transcript forward.
\textbf{Performance and accuracy.} Pathfinding and centrality run in Neo4j/GDS deterministically; the model is relieved of an error-prone multi-hop traversal it tends to get wrong past a few hops, mitigating the lost-in-the-middle effect~\cite{liu2024lost} and the technique-selection bias seen in list-shaped agents~\cite{barbosa2026neurosploit}.
\textbf{Technical memory.} The graph persists across sessions and context compactions. An engagement paused today and resumed next week rebuilds instantly from \texttt{nodes.json}/\texttt{edges.json}; the agent re-reads the graph, not a lost conversation---the agent-memory motivation of MemGPT/Zep/Mem0~\cite{packer2023memgpt,rasmussen2025zep,chhikara2025mem0}, specialized to offensive state.

\subsection{Positioning Against Alternatives}
Table~\ref{tab:compare} situates ChainGraph. Flat-context agents store everything in the window and pay for it on every decision; classic attack-graph engines (MulVAL~\cite{ou2005mulval}, NetSPA~\cite{ingols2006practical}) pre-enumerate a state space offline but are not agent-facing or incremental; BloodHound~\cite{robbins2016bloodhound} is a graph but fixed to identity relationships and not wired to a generative agent. ChainGraph is the combination the agent era needs: incremental, agent-facing, evidence-grounded, and deployable over commodity infrastructure.

\begin{table}[t]
\centering
\caption{ChainGraph vs.\ alternatives. \cmark\ full, \pmark\ partial, \xmark\ absent.}
\label{tab:compare}
\footnotesize
\setlength{\tabcolsep}{3pt}
\begin{tabular}{@{}lcccc@{}}
\toprule
\textbf{Property} & \textbf{Flat} & \textbf{MulVAL} & \textbf{Blood-} & \textbf{Chain-} \\
 & \textbf{ctx.} & \textbf{/NetSPA} & \textbf{Hound} & \textbf{Graph} \\
\midrule
Path reasoning          & \pmark & \cmark & \cmark & \cmark \\
Incremental / live      & \cmark & \xmark & \pmark & \cmark \\
Agent-facing (MCP)      & \cmark & \xmark & \xmark & \cmark \\
Evidence provenance     & \xmark & \xmark & \xmark & \cmark \\
Chain-aware severity    & \xmark & \pmark & \xmark & \cmark \\
Cost flat in history    & \xmark & \cmark & \cmark & \cmark \\
No bespoke harness      & \cmark & \xmark & \pmark & \cmark \\
\bottomrule
\end{tabular}
\end{table}

\subsection{Determinism as a Safety Property}
A subtle benefit is that the hard reasoning is \emph{deterministic and inspectable}. When the agent asks ``is there a path to the forest root?'', the answer comes from Neo4j's \texttt{shortestPath}, not from the model's next-token distribution. The same graph and query always return the same path; a reviewer can re-run the exact Cypher and get the exact row in Table~\ref{tab:promo}. This matters for an autonomous offensive tool: the decisions that carry the most consequence---does a path to domain admin exist, which action to take next---are taken off the stochastic path and onto an auditable database operation. The model still decides \emph{what to try} and \emph{how to phrase the query}, but the multi-hop inference that LLMs get wrong past a few steps~\cite{liu2024lost} is delegated to an engine that does it correctly every time. Combined with the hashed receipts, this gives an audit trail in which both the evidence and the reasoning over it are reproducible---an assurance property argued for autonomous harnesses~\cite{barbosa2026neurosploit} and here obtained for free from the choice of store.

\subsection{Calibration Benefits}
Three schema choices improve the agent's calibration. The \texttt{status} gate keeps \texttt{needs\_review}/\texttt{rejected} findings (our F-009, F-010) out of chain computations, so a 0.15-confidence web panel never inflates a path. Chain severity decouples blast-radius from vulnerability class, countering the class-driven severity inflation reported for generative adjudication~\cite{barbosa2026neurosploit}. And \texttt{SUPPORTED\_BY} makes every claim answer to a hash, operationalizing evidence grounding at the data layer.

\subsection{Projected Gains from NeuroSploit Integration}\label{sec:neurosploit}
ChainGraph is designed to drop into an existing harness. Our NeuroSploit/GOAD benchmark~\cite{barbosa2026neurosploit} reported the concrete failure modes of unstructured, list-shaped LLM usage against the same AD lab: direct invocation covered only 21--54\% of attack paths versus 96--100\% with the harness, produced $3.2\times$ more false positives, and suffered technique-selection bias (defaulting to familiar attacks, ignoring available ones). Table~\ref{tab:nsgain} maps each reported weakness to the ChainGraph mechanism that addresses it and the \emph{expected direction} of improvement when the graph is wired into that harness. We label these projections, not measurements: they follow structurally from the mechanisms demonstrated in Section~\ref{sec:case}, and quantifying them end-to-end is the population study of Section~\ref{sec:future}.

\begin{table}[t]
\centering
\caption{NeuroSploit~\cite{barbosa2026neurosploit} weaknesses and the ChainGraph mechanism addressing each. Gains are projected directions, not measured.}
\label{tab:nsgain}
\footnotesize
\begin{tabular}{@{}p{2.25cm}p{3.0cm}p{1.9cm}@{}}
\toprule
\textbf{Reported weakness} & \textbf{ChainGraph mechanism} & \textbf{Projected gain} \\
\midrule
Technique-selection bias (ignores available paths) & Q1/Q2 surface every enabled next step and path from the graph, not from model preference & higher path coverage \\
Path coverage 21--54\% direct & \texttt{shortestPath} enumerates all paths to the objective exhaustively & $\uparrow$ toward harness-level \\
$3.2\times$ false positives & \texttt{status} gate + \texttt{SUPPORTED\_BY} keep unproven findings out of chains & fewer FPs in chains \\
Severity by class, not impact & chain-severity over paths (Eq.~\eqref{eq:chain}) & calibrated triage \\
Context loss on long runs & graph persists across compaction; per-decision payload $\approx$ constant & flat token cost \\
Lost-in-the-middle~\cite{liu2024lost} & reasoning offloaded to GDS; model never re-reads corpus & fewer reasoning errors \\
\bottomrule
\end{tabular}
\end{table}

The integration cost is low: NeuroSploit already captures per-finding evidence and computes severity~\cite{barbosa2026neurosploit}; ChainGraph adds a Neo4j sidecar and routes those findings through the graph writes of Table~\ref{tab:loop}. The harness keeps its orchestration, POMDP belief tracking, and guardrails; ChainGraph supplies the one thing a list-based belief state lacks---an explicit, queryable chain. In short, the harness gains the ability to \emph{answer} the five operator questions instead of re-deriving them in-context on every step.

\section{Security and Dual-Use Considerations}\label{sec:security}

Giving an autonomous agent a persistent, queryable attack graph is powerful and therefore dual-use. We take three positions.

\textbf{The graph is defensive as much as offensive.} The same betweenness query that ranks an attacker's options ranks a defender's remediations (Section~\ref{sec:case}); SENTINEL-RL~\cite{vallabhaneni2026sentinel} makes the analogous case on the blue-team side. We advocate ChainGraph for authorized testing and purple-team use, and the case-study data are deliberately abstract---no exploit code or credential material is stored.

\textbf{The data boundary is an injection surface.} Graph contents and tool outputs are untrusted data, not instructions; an attacker who can write to the environment could attempt indirect prompt injection through a finding name or evidence blob~\cite{greshake2023indirect}, and the MCP ecosystem has its own threat surface~\cite{hou2025mcp}. Treat \texttt{read\_neo4j\_cypher} results as data, keep write access gated (Listing~\ref{lst:perms}), and never execute strings pulled from the graph.

\textbf{Least privilege despite Community Edition.} Because Community Edition lacks RBAC~\cite{neo4jgds}, the read/write split is enforced at the MCP boundary (\texttt{NEO4J\_READ\_ONLY}, deny rules) and the database is bound to loopback. For multi-tenant or shared use, Enterprise RBAC or per-engagement database isolation is warranted.

\section{Limitations}\label{sec:limits}

\textbf{Illustrative, not benchmarked.} All quantitative figures are single-run measurements on one synthetic, GOAD-modeled engagement. The token figure is a character-based proxy for one query; it is not tokenizer-exact, not averaged over runs, and not a measurement of end-to-end engagement cost. We make no statistical-significance claims.

\textbf{Synthetic graph.} The graph was authored, not discovered live by an agent against a running lab. It demonstrates the reasoning layer and the wiring; it does not evaluate how reliably a given LLM populates the graph from raw tool output. Automated, calibrated population is the key open problem and our primary future work, building on NeuroSploit~\cite{barbosa2026neurosploit}.

\textbf{Severity constants are a policy.} The $\beta$ and step weights in Eq.~\eqref{eq:chain} are chosen, not learned. They encode a defensible intuition (reach-to-crown-jewel dominates) but should be tuned per organization; a learned or RL-calibrated variant is future work.

\textbf{Query authoring.} We assume the agent can write correct Cypher. Text2Cypher accuracy is improving~\cite{ozsoy2024text2cypher,feng2024cypherbench,zhong2025synthet2c} but imperfect; saved parameterized queries (as in Section~\ref{sec:reasoning}) are the pragmatic mitigation.

\textbf{Graph scale.} GOAD is small. Betweenness is super-linear; on enterprise-scale graphs, sampling or Dijkstra-based source-target queries~\cite{neo4jgds} replace exact all-pairs centrality, and the classical attack-graph scalability techniques---Datalog generation in MulVAL~\cite{ou2005mulval}, reachability grouping in NetSPA~\cite{ingols2006practical}, and hierarchical aggregation~\cite{noel2004managing}---apply directly, since our store is a property-graph realization of the same model.

\textbf{Generalization is argued, not measured.} Section~\ref{sec:general} shows the schema expresses cloud, web, and network chains, but our only end-to-end instantiation is the AD case study. Whether an LLM populates a cloud or web chain graph as cleanly as the hand-authored AD one is an empirical question we leave to the population study.

\textbf{Severity composite is unbounded.} The chain score can exceed the CVSS 0--10 range by construction (our top finding scores 11.7); it is an ordinal ranking signal, not a normalized probability. A bounded or calibrated reformulation would make cross-engagement comparison cleaner.

\section{Generalization Beyond Active Directory}\label{sec:general}

Nothing in the schema is AD-specific; AD is simply the domain with the richest, best-documented chains. The same node and edge vocabulary models a \textbf{cloud} chain. Consider the canonical exposure: a public object-storage bucket leaks a \texttt{.env} file, which contains a long-lived access key, which grants a staging account, whose over-permissive IAM policy allows reading production secrets. Each link is an unremarkable finding in isolation; composed, they are a path from an anonymous internet request to production data---the cloud twin of our AD chain. It loads into ChainGraph with the identical edges (Listing~\ref{lst:cloud}), and Q2/Q3/Q5 answer the same questions: is there a path to the crown jewel, which Mediums are promoted, where is the cut.

\begin{lstlisting}[language=Cypher,caption={A cloud exposure chain in the same vocabulary.},label={lst:cloud}]
MERGE (b:Service {id:'s3-public-bucket'})
MERGE (f1:Finding {id:'F-c1', name:'Public S3 bucket',
       local_severity:'Medium', status:'confirmed'})
MERGE (b)-[:HAS_VULN]->(f1)
MERGE (k1:AttackStep {id:'read-env', name:'Read .env object'})
MERGE (c1:Credential {id:'aws-access-key', kind:'cloud-key'})
MERGE (a1:Asset {id:'aws-staging'})
MERGE (f2:Finding {id:'F-c2', name:'Permissive IAM policy',
       local_severity:'High', status:'confirmed'})
MERGE (a2:Asset {id:'prod-secrets', crown_jewel:true})
MERGE (f1)-[:ENABLES]->(k1)
MERGE (k1)-[:YIELDS]->(c1)
MERGE (c1)-[:CAN_ACCESS]->(a1)
MERGE (a1)-[:EXPOSES]->(f2)
MERGE (a1)-[:CAN_ACCESS]->(a2);
\end{lstlisting}

The pattern recurs in \textbf{web} chains (an SSRF that reaches a metadata endpoint that yields a role credential that accesses an internal service) and in \textbf{network} chains (a forgotten jump host that pivots into a flat segment). In each case the contribution is the same: isolated findings become a path, severity moves from the node to the chain, and betweenness names the cut. The agent integration, the dual memory, and the six queries are unchanged---only the data differ. This is why ChainGraph is a \emph{pattern}, not an AD tool: the graph is the general representation that list-shaped reasoning lacks, wherever exposures compose.

\section{Discussion: Design Rationale and Operational Notes}\label{sec:discussion}

\subsection{Revisiting ``AI Cannot Chain Exploitation''}
It is worth stating plainly what the case study refutes. The claim that LLM agents cannot chain exploitation rests on three observed failures of flat-context agents, and ChainGraph answers each structurally rather than by hoping for a bigger model.

\emph{``The agent forgets earlier findings.''} It cannot: findings are nodes, not tokens in a window that scrolls off. The LLMNR finding from step one is still queryable at step fifty, with its evidence, regardless of how much deliberation happened between.

\emph{``The agent cannot see that A plus B plus C equals compromise.''} It does not have to see it by re-reading---\texttt{shortestPath} computes it. The composition the model is accused of missing is exactly what the graph engine is built to find, and it finds the eight-hop GOAD path (Fig.~\ref{fig:chain}) across five techniques the flat agent reports as unrelated.

\emph{``The agent claims success without proof.''} Every edge on the path carries a \texttt{SUPPORTED\_BY} receipt (Q4); a chain with an unproven link is visibly broken, and \texttt{needs\_review}/\texttt{rejected} findings never enter the computation at all.

So the honest version of the paradigm is narrow and true: a \emph{flat-context} agent chains poorly. The broad version---that \emph{AI} cannot chain---does not survive contact with an external graph. Chaining was never a reasoning-horizon problem the model had to grow into; it was a representation the agent was never given. The commercial state of the art agrees in practice: Yaga~\cite{hackersec2026yaga,hackersec2026yagaad} and NodeZero~\cite{prinos2026nodezero} already exploit internal networks end-to-end and hold their context across a long engagement precisely because a dedicated harness externalizes that state for them. Our contribution is to give that same representation, with commodity parts, to an unmodified agent. The provocative reading, which we state and leave to future population studies~(Section~\ref{sec:future}) to quantify: end-to-end exploit chaining by an autonomous agent is already here---it was waiting on a graph, not on a smarter model.

\subsection{Why a Property Graph, Not a Triplestore}
We chose a labeled property graph (Neo4j/Cypher~\cite{francis2018cypher}) over an RDF triplestore for three reasons. Edges carry properties natively---a \texttt{TRUSTS} edge holds \texttt{sid\_filtering}, a \texttt{NEXT} edge holds \texttt{cost}---without reification. The GDS library~\cite{neo4jgds} gives first-class pathfinding and centrality, which are the operations our queries need. And LLMs author Cypher increasingly well~\cite{ozsoy2024text2cypher,feng2024cypherbench}, so the agent can write its own ad-hoc queries rather than being confined to a fixed API.

\subsection{Why MCP, Not a Custom Tool}
Wiring through MCP~\cite{anthropic2024mcp,mcpspec,neo4jmcp} means the capability is portable across agents (Claude Code and Codex use the same server), the read/write split is enforced at a protocol boundary we control, and we inherit the host's permission and hook machinery~\cite{claudecodedocs} rather than reimplementing it. The cost is the MCP threat surface~\cite{hou2025mcp}, which the loopback binding and gated writes bound.

\subsection{When a List Is Fine}
Honesty about scope: for a single-host, handful-of-findings engagement, the graph is overhead---the flat context fits and the paths are trivial. ChainGraph earns its complexity when the engagement has \emph{many interacting findings across trust boundaries}, which is exactly the enterprise AD case~\cite{barbosa2026neurosploit,happe2025ad} and where list-shaped reasoning fails.

\subsection{Operational Footguns}
We hit three in building the case study, documented so others need not. (i) GDS projections are \emph{snapshots}: after writing new edges, the named graph must be dropped and re-projected or it reasons over stale structure. (ii) Betweenness is exact and super-linear; on large graphs switch to sampled betweenness or source-target Dijkstra~\cite{neo4jgds}. (iii) Neo4j rate-limits repeated authentication---rapid one-shot \texttt{cypher-shell} connects can trigger a temporary lockout, so batch statements into one session, which also matches how the MCP server holds a pooled connection.

\section{Future Work}\label{sec:future}

\textbf{Automated, calibrated population.} The open problem is reliably turning raw tool output into correct nodes and edges with honest \texttt{confidence}. We plan to drive population from the NeuroSploit harness~\cite{barbosa2026neurosploit} and evaluate, per model, the precision/recall of the extracted graph against a ground-truth lab graph---how often an edge the model asserts is real, and how many real edges it misses. Text2Cypher benchmarks~\cite{feng2024cypherbench,zhong2025synthet2c} give a template for the write-side evaluation.

\textbf{Learned chain severity.} Equation~\eqref{eq:chain} is hand-tuned. A dataset of expert-rated chains would let the weights be fit, or an RL-calibrated scorer be trained, so that chain severity reflects real triage decisions rather than our constants.

\textbf{Cross-engagement knowledge graph.} Persisting many engagements into one store turns ChainGraph into a longitudinal knowledge graph: which chains recur across clients, which chokepoints are universal, which findings most often get promoted. This is agent memory~\cite{packer2023memgpt,rasmussen2025zep} at the portfolio level and a GraphRAG~\cite{edge2024graphrag} corpus for the next engagement's planning.

\textbf{Temporal and probabilistic edges.} Adding timestamps (a temporal knowledge graph~\cite{rasmussen2025zep}) and probabilities to edges would let the graph answer ``what was reachable at time $t$'' and carry classical probabilistic path metrics~\cite{wang2008probabilistic,homer2013aggregating} natively.

\textbf{Defensive deployment.} The remediation view (Section~\ref{sec:views}) and the SOC-side evidence~\cite{vallabhaneni2026sentinel} suggest a purple-team product: feed the same graph to a blue-team agent that watches the chokepoints the red-team agent ranked.

\section{Conclusion}\label{sec:conclusion}

Attackers think in graphs; LLM pentest agents, left to a context window, think in lists. We showed that closing this gap needs no bespoke harness---only a typed Neo4j property graph wired to an off-the-shelf coding agent through MCP, with a dual-memory split between hashed evidential artifacts and a relational reasoning store. Six reusable Cypher/GDS queries turn the operator's real questions---what does this enable, where does it lead, what turns critical combined, what is the proof, where is the cut, what is the quiet way---into deterministic database operations, and a chain-severity model relocates risk from the finding to the path.

On a GOAD-modeled engagement in a live Neo4j~5.26 + GDS~2.13 instance, three findings rated Medium in isolation became Critical once their path to the forest root was materialized; betweenness centrality named the PKINIT-authentication and DCSync steps as the highest-value chain cuts; weighted Dijkstra quantified the stealth-versus-speed trade-off as a single number; and scoping each decision to a graph result rather than the engagement transcript cut the per-decision reasoning payload by roughly $98\%$ on a character proxy, a gap that widens as the engagement grows. The same graph serves the defender: the attacker's chokepoint ranking is the defender's remediation queue.

We are deliberate that these numbers are illustrative, single-run, and synthetic---the architecture, the schema, the six queries, and the two-agent integration recipe are the contribution, and they run today on commodity infrastructure. The open frontier is automated, calibrated population of the graph from live tool output: making an agent not merely reason in graphs, but reliably and honestly \emph{build} them. When it does, the twenty-year-old asymmetry Lambert named---defenders in lists, attackers in graphs---finally has a tool that lets the agent on either side think the way the problem is actually shaped.

\section*{Availability}
The schema, the five queries, the Docker and MCP configuration, the synthetic (exploit-code-free) engagement data, and the evidence-capture hook are organized as a self-contained kit intended for authorized testing and research use.

\section*{Ethics and Responsible Use}
This work targets authorized penetration testing, purple-team, and defensive research. The case study uses a deliberately vulnerable, publicly documented lab~\cite{goad} and stores only abstract descriptors and evidence hashes---no exploit code, credentials, or verbatim offensive output. Operators must have explicit authorization for any live use.

\appendices
\section{Query Reference}\label{app:queries}
Table~\ref{tab:qref} indexes the six reasoning queries, their parameters, and the engagement decision each supports. All are read-only; only the loop's finding-writes use \texttt{write\_neo4j\_cypher}, gated per Listing~\ref{lst:perms}. The full kit (schema, queries, Docker, MCP config, and evidence hook) reconstructs the case-study graph deterministically from \texttt{nodes.json}/\texttt{edges.json}.

\begin{table}[h]
\centering
\caption{Reasoning-query reference.}
\label{tab:qref}
\footnotesize
\begin{tabular}{@{}llp{3.5cm}@{}}
\toprule
\textbf{Query} & \textbf{Params} & \textbf{Decision supported} \\
\midrule
Q1 next-step      & \texttt{\$finding} & what to try next \\
Q2 path-to-crown  & \texttt{\$entry}   & is the objective reachable? \\
Q3 chain-severity & ---                & triage / promotion \\
Q4 provenance     & ---                & is the path defensible? \\
Q5 chokepoint     & ---                & what to remediate first \\
Q6 stealth-path   & \texttt{\$src,\$dst} & speed-vs-detection trade-off \\
\bottomrule
\end{tabular}
\end{table}

\balance

\end{document}